\documentclass[final, a4paper, 12pt]{article}

\usepackage[utf8]{inputenc}
\newcommand{\poo}[2]{\frac{\partial #1}{\partial #2}}
\newcommand{\po}[3]{\frac{\partial^{#3}#1}{\partial #2^{#3}}}
\newcommand{\p}{\partial}

\usepackage{amsmath}
\usepackage{graphicx}
\usepackage{amssymb}
\usepackage{mathrsfs}  
\usepackage{subfigure}
\usepackage{caption}
\usepackage{epstopdf}
\usepackage{authblk}

\usepackage[sort&compress, comma, square, numbers]{natbib}

\usepackage{a4wide}

\usepackage[dvipsnames, table]{xcolor}
\definecolor{bckg}{RGB}{20.8, 20.8, 20.8}
\definecolor{oneblue}{rgb}{0.0, 0.0, 0.85}
\definecolor{Lightblue}{RGB}{214, 214, 214}
\definecolor{bluepigment}{rgb}{0.2, 0.2, 0.6}
\definecolor{charcoal}{rgb}{0.21, 0.27, 0.31}
\definecolor{denimblue}{rgb}{0.08, 0.38, 0.74}
\definecolor{Lightgray}{rgb}{0.89, 0.89, 0.89}
\definecolor{darkgrey}{rgb}{0.273, 0.281, 0.30}
\definecolor{darkelectricblue}{rgb}{0.33, 0.41, 0.47}

\usepackage[colorlinks,
           urlcolor=oneblue,
           linkcolor=denimblue,
           citecolor=NavyBlue,
           bookmarksopen=false,
           pdfpagemode=UseNone,
           pagebackref]{hyperref}

\usepackage{dsfont}
\newcommand{\R}{\mathds{R}}
\newcommand{\ud}{\mathrm{d}}
\numberwithin{equation}{section}

\title{Free Surface Flows in Electrohydrodynamics with a Constant Vorticity Distribution}

\author{M. J. Hunt \\ \textsf{Warwick Manufacturing Group, University of Warwick, Coventry CV4 7AL, UK and Warwick Mathematics Institute, Zeeman Building, University of Warwick, Coventry CV4 7AL, UK} \and D. Dutykh \\ \textsf{Univ. Grenoble Alpes, Univ. Savoie Mont Blanc, CNRS, LAMA, 73000 Chamb\'ery, France.}}

\date{\today}

\begin{document}

\maketitle

\begin{abstract}
In 1895, Korteweg and de Vries (KdV), \cite{kdv}, derived their celebrated equation describing the motion of waves of long wavelength in shallow water. In doing so they made a number of quite reasonable assumptions, incompressibility of the water and irrotational fluid. The resulting equation, the celebrated KdV equation, has been shown to be a very reasonable description of real water waves. However there are other phenomena which have an impact on the shape of the wave, that of vorticity and viscosity. This paper examines how a constant vorticity affects the shape of waves in electrohydrodynamics. For constant vorticity, the vertical component of the velocity obeys a Laplace equation and also has the usual lower boundary condition. In making the vertical component of the velocity take central stage, the Burns condition can be thus bypassed.
\end{abstract}


\section{Introduction}

Water waves constitute a very classical problem in hydrodynamics \cite{Craik2004}. However, this problem is traditionally formulated in terms of the velocity potential to achieve some simplifications. In other words, there has always been an implicit assumption of zero vorticity in the flow region. In numerous recent studies this assumption started to be questioned. One of pioneering studies was made by Burns (1953) \cite{Burns1953}. Later, Da~Silva and Peregrine (1988) \cite{DaSilva1988} studied steep and steady waves on finite depth with constant vorticity. A constant distribution is the next logical step after identically zero distribution. More recently, this problem was analyzed mathematically in some two-component systems \cite{Escher2016}. The effect of the vorticity on travelling wave solutions (solitary and cnoidal) was investigated in the purely hydrodynamic context in \cite{Dutykh2019} using the qualitative phase space analysis methods. A Hamiltonian formulation has been reported in \cite{Wahlen2007}. However, this problem in electrohydrodynamics seems to be still open to the best of our knowledge. The present study should be considered as a further attempt to fill in this gap in the literature.

The current approach to examining flows with constant vorticity in two dimensions is via the use of a stream function, $\psi$, and it's harmonic conjugate, the velocity potential, $\varphi$, so $\mathbf{u}=\nabla\varphi+\nabla^{\perp}\psi$. This approach is introduces two essentially unnecessary functions which complicates the problem and has the limiting effect in being restricted to fully nonlinear and linear computations. There has been no attempt to undertake a weakly nonlinear analysis which is the purpose of this manuscript. Previous work on constant vorticity models typically have been fully nonlinear, for example see \cite{jmvb2000}.

The present manuscript is organized as follows. The problem is formulated in Section~\ref{sec:for}. The linear analysis of this problem is performed in Section~\ref{sec:lin}, while the weakly nonlinear analysis is presented in Section~\ref{sec:wn}. Some numerical predictions of the weakly nonlinear theory are presented in Section~\ref{sec:num}. Finally, the main conclusions and perspectives of this study are outlined in section~\ref{sec:concl}.

\section{Formulation}
\label{sec:for}

\begin{figure}
\begin{center}
\includegraphics[width=0.75\textwidth]{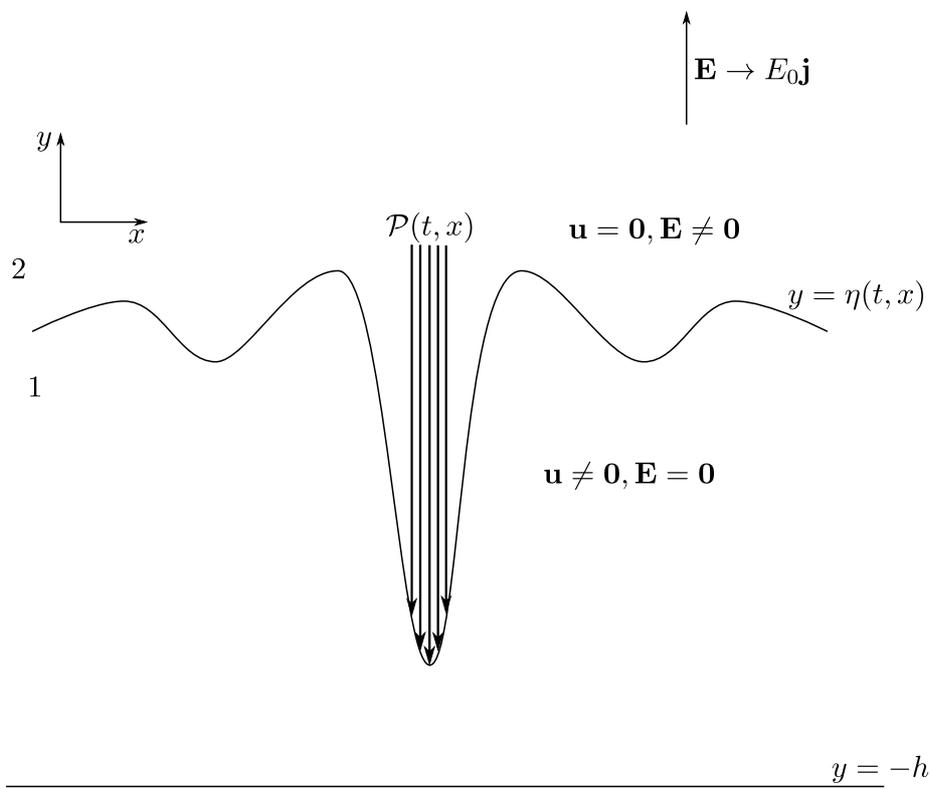}
\caption{\small\em Physical problem schematic representation.}
\label{fig1}
\end{center}
\end{figure}

A two dimensional fluid in region $1$ is considered which is incompressible and inviscid. The vorticity, $\omega$ is constant as is the surface tension $\sigma$. Cartesian co-ordinates are introduced as shown in figure \ref{fig1}. Region $1$ is defined as $-h<y<\eta(t,x)$ $\forall x\in\mathbb{R}$. The moving pressure distribution $\mathcal{P}(t,x)$ is chosen to act along the interface $y=\eta (t,x)$ and $\mathcal{P}\rightarrow 0$ as $|x|\rightarrow 0$.

In region $2$ defined by $\left\lbrace (x,y)|x\in\mathbb{R},y>\eta (t,x)\right\rbrace$ there is an electric field, $\mathbf{E}$ which has no charges and is therefore obtained by a potential $\mathbf{E}=-\nabla V$. The potential is chosen as $V(x,-h)=0$ and as the fluid is perfectly conducting this also means that $V(x,\eta (t,x))=0$. A vertical electric field is set up by imposing:
\begin{equation}
V\sim -E_{0}y,\quad y\rightarrow\infty
\end{equation}
The equation for the electric potential is therefore given by:
\begin{equation}
\po{V}{x}{2}+\po{V}{y}{2}=0,
\end{equation}
with the condition:
\begin{equation}
\left[\mathbf{E}\cdot\hat{\mathbf{t}}\right]_{1}^{2}=0
\end{equation}
which is one of the general boundary conditions derivable from Maxwell's equations. In region $1$, the Navier-Stokes equations are used with the stress tensor:
\begin{equation}
T_{ij}=-P\delta_{ij}+\Sigma_{ij}
\end{equation}
Where:
\begin{equation}
\Sigma_{ij}=\epsilon_{p}\left(\poo{V}{x_{i}}\poo{V}{x_{j}}-\frac{1}{2}\delta_{ij}\sum_{k=1}^{2}\poo{V}{x_{k}}\poo{V}{x_{k}}\right).
\end{equation}
Where $\epsilon_{p}$ is called the electric permittivity. The tensor $\Sigma_{ij}$ has various names, in the fluids literature it is known as the Maxwell-stress tensor. It can be shown that:
\begin{equation}
\sum_{i=1}^{2}\poo{\Sigma_{ij}}{x_{i}}=0.
\end{equation}
So The Navier-Stokes equations reduce to the Euler equations.
\begin{equation}
\poo{\mathbf{u}}{t}+(\mathbf{u}\cdot\nabla )\mathbf{u}=-\frac{1}{\rho}\nabla P -g\mathbf{j},
\end{equation} 
where $\mathbf{u}=(u,v)$ is the velocity and $P$ is the pressure in the fluid.
The boundary $y=-h$ is taken as impenetrable, so $v(x,-h)=0$. The fluid is incompressible and has constant vorticity, $\omega$ and so:
\begin{equation}
\poo{u}{x}+\poo{v}{y}=0,\quad \poo{v}{x}-\poo{u}{y}=\omega .
\end{equation}   
These equations can be cross differentiated to obtain a single equation for $v$,
\begin{equation}\label{gov}
\po{v}{x}{2}+\po{v}{y}{2}=0.
\end{equation}
The free surface equation is given by:
\begin{equation}
\poo{\eta}{t}+u\poo{\eta}{x}=v, \quad y=\eta (t,x).
\end{equation}
This gives a boundary condition for $v$ on $y=\eta (t,x)$. The lower boundary for $v$ is given by:
\begin{equation}\label{lbdy}
    v(t,x,-h)=0
\end{equation}
Equations (\ref{gov})-(\ref{lbdy}) are the core of the technique in deriving the free surface profiles. The other boundary condition used is the Young-Laplace equation given by:
\begin{equation}
\left[\hat{\mathbf{n}}\cdot\mathbf{T}\cdot\hat{\mathbf{n}}\right]_{1}^{2} =\sigma\frac{\p_{x}^{2}\eta}{(1+(\p_{x}\eta )^{2})^{\frac{3}{2}}}
\end{equation}
The Euler equations may be simplified using electrostatic equilibrium. The equilibrium condition is:
\begin{equation}
-\frac{1}{\rho}\poo{p}{y}-g=0
\end{equation}
Integrating this equation shows that $p=-\rho gy+C$ To compute the value of $C$, use the Young-Laplace equation to see $C=P_{a}-\epsilon_{d}E_{0}^{2}/2$, here $P_{a}$ is the atmospheric pressure. So now write:
\begin{equation}
P=P_{a}-\frac{\epsilon_{p}E_{0}^{2}}{2}-\rho gy+p
\end{equation}
The Euler equations now become:
\begin{equation}
\poo{\mathbf{u}}{t}+(\mathbf{u}\cdot\nabla )\mathbf{u}=-\frac{1}{\rho}\nabla p
\end{equation}
The Young-Laplace equation becomes:
\begin{equation}
p-\rho g\eta -\frac{\epsilon_{p}E_{0}^{2}}{2}=\mathcal{P}-\frac{1}{1+(\p_{x}\eta )^{2}}\left[(\p_{x}\eta )^{2}\Sigma_{11}-2\p_{x}\eta\Sigma_{12} +\Sigma_{22}\right] -\sigma\frac{\p_{x}^{2}\eta}{(1+(\p_{x}\eta )^{2})^{\frac{3}{2}}}
\end{equation}

\section{Linear Theory}
\label{sec:lin}

The scaling for the linear theory is:
\begin{equation}
(x,y,\eta )=h(\hat{x},\hat{y},\hat{\eta}),\quad t=\sqrt{\frac{\rho h^{3}}{\sigma}}\hat{t},\quad (p,\mathcal{P}) =\frac{\sigma}{h}(\hat{p},\hat{\mathcal{P}}), \quad \mathbf{u}=\sqrt{\frac{\sigma}{\rho h}}\hat{\mathbf{u}}
\end{equation}
\begin{equation}
\mathcal{P}=\rho gh\hat{\mathcal{P}}\quad V=hE_{0}\hat{V}
\end{equation}
The equations become:
\begin{equation}
\po{\hat{V}}{\hat{x}}{2}+\po{\hat{V}}{\hat{y}}{2}=0,\quad \hat{y}>\hat{\eta}
\end{equation}
\begin{equation}
\poo{\hat{V}}{\hat{x}}+\poo{\hat{\eta}}{\hat{x}}\poo{\hat{V}}{\hat{y}}=0,\quad \hat{y}=\hat{\eta}
\end{equation}
\begin{equation}
\poo{\hat{\mathbf{u}}}{\hat{t}}+(\hat{\mathbf{u}}\cdot\hat{\nabla})\hat{\mathbf{u}}=-\hat{\nabla}\hat{p},\quad -1<\hat{y}<\hat{\eta}
\end{equation}
\begin{equation}
\poo{\hat{u}}{\hat{x}}+\poo{\hat{v}}{\hat{y}}=0,\quad\poo{\hat{v}}{\hat{x}}-\poo{\hat{u}}{\hat{y}}=\Omega ,\quad -1<\hat{y}<\hat{\eta}
\end{equation}
\begin{equation}
\poo{\hat{\eta}}{\hat{t}}+\hat{u}\poo{\hat{\eta}}{\hat{x}}=\hat{v},\quad \hat{y}=\hat{\eta}
\end{equation}
\begin{equation}
\po{\hat{v}}{\hat{x}}{2}+\po{\hat{v}}{\hat{y}}{2}=0
\end{equation}
\begin{multline}
\hat{p}-B\hat{\eta}-\frac{E_{b}}{2}=\hat{\mathcal{P}}-\frac{E_{b}}{1+(\p_{\hat{x}}\hat{\eta})^{2}}\left[ (\p_{\hat{x}}\hat{\eta})^{2}\hat{\Sigma}_{11}-2\p_{\hat{x}}\hat{\eta}\hat{\Sigma}_{12}+\hat{\Sigma}_{22}\right] - \\
-\frac{\p_{\hat{x}}^{2}\hat{\eta}}{(1+(\p_{\hat{x}}\hat{\eta})^{2})^{\frac{3}{2}}}, \quad \hat{y}=\hat{\eta}
\end{multline}
Where:
\begin{equation}
\Omega =\omega\sqrt{\frac{\rho h^{3}}{\sigma}},\quad B=\frac{\rho gh^{2}}{\sigma},\quad E_{b}=\frac{\epsilon_{p}E_{0}^{2}}{\sigma}
\end{equation}
Expand according to:
\begin{eqnarray}
\hat{u} & = & -\Omega \hat{y}+\varepsilon u_{1}+o(\varepsilon ) \\
\hat{v} & = & \varepsilon v_{1}+o(\varepsilon ) \\
\hat{p} & = & \varepsilon p_{1}+o(\varepsilon ) \\
\hat{\mathcal{P}} & = & \varepsilon\mathcal{P}_{1}+o(\varepsilon ) \\
\hat{\eta} & = & \varepsilon\eta_{1}+o(\varepsilon ) \\
\hat{V} & = & -\hat{y}+\varepsilon V_{1}+o(\varepsilon )
\end{eqnarray}
The set of linear equations now becomes:
\begin{eqnarray}
\po{V_{1}}{\hat{x}}{2}+\po{V_{1}}{\hat{y}}{2} & = & 0,\quad \hat{y}>0, \\
\poo{V_{1}}{\hat{x}}-\poo{\eta_{1}}{\hat{x}} & = & 0, \quad \hat{y}=0,\label{disp_1} \\
\poo{u_{1}}{\hat{t}}-\Omega y\poo{u_{1}}{\hat{x}}-\Omega v_{1} & = & -\poo{p_{1}}{\hat{x}} ,\quad -1<\hat{y}<0\label{disp_3} \\
\poo{v_{1}}{\hat{t}}-\Omega y\poo{v_{1}}{\hat{x}} & = & -\poo{p_{1}}{y},\quad -1<\hat{y}<0 \\
\po{v_{1}}{\hat{x}}{2}+\po{v_{1}}{\hat{y}}{2} & = & 0,\quad -1<\hat{y}<0 \\
\poo{\eta_{1}}{\hat{t}} & = & v_{1},\quad \hat{y}=0,\label{disp_2} \\
p_{1}-B\eta_{1} & = & \mathcal{P}_{1}+E_{b}\poo{V_{1}}{\hat{y}}-\po{\eta_{1}}{\hat{x}}{2}
\end{eqnarray} 
\subsection{General Dispersion Relation}
To obtain a dispersion relation set $\mathcal{P}_{1}=0$ and write all perturbations in the form:
\begin{equation}
f(t,x,y)=\frac{1}{2\pi}\int_{\mathbb{R}}\hat{f}(\xi (k),k,y)e^{i(kx-\xi t)}dk
\end{equation}
Solving the equation for $v_{1}$ and $V_{1}$ shows that:
\begin{equation}
v_{1}=\alpha\sinh k(y+1),\quad V_{1}=\beta e^{-|k|y}
\end{equation}
Using equations (\ref{disp_1}) and (\ref{disp_2}) yields $\beta$ in terms of $\alpha$:
\begin{eqnarray}
\beta =-\frac{i}{\xi}\alpha\sinh k
\end{eqnarray}
Only $p_{1}$ at the surface is required, so it is possible to set $y=0$ in equation (\ref{disp_3}) to find:
\begin{eqnarray}
\hat{p}_{1}=\frac{\alpha i}{k}(\xi \cosh k -\Omega\sinh k)
\end{eqnarray}
Inserting everything into the linearised Young-Laplace equations shows:
\begin{equation}
\xi^{2}-\xi\Omega\tanh k-Bk\tanh k+E_{b}k|k|\tanh k-k^{3}\tanh k=0
\end{equation}
The phase velocity, $c$ is given by $c=\xi /k$ and solving for the phase velocity shows that:
\begin{equation}\label{disp_rel}
c=\frac{\Omega\tanh k}{2k}\pm\frac{1}{2k}\sqrt{\Omega^{2}\tanh^{2}k+4(Bk-E_{b}k|k|+k^{3})\tanh k}
\end{equation}
It can be seen that setting $\Omega =0$ reduces to the dispersion relation in \cite{Hunt1} and setting $E_{b}=0$ results in the dispersion relation in \cite{Hur}. It was noted in \cite{Hunt1} that in order to have a linear wave profile the parameters had to satisfy the inequality $4B\geqslant E_{b}^{2}$, however with the inclusion of positive vorticity this is no longer the case. 

\begin{figure}
\begin{center}
\includegraphics[width=0.79\textwidth]{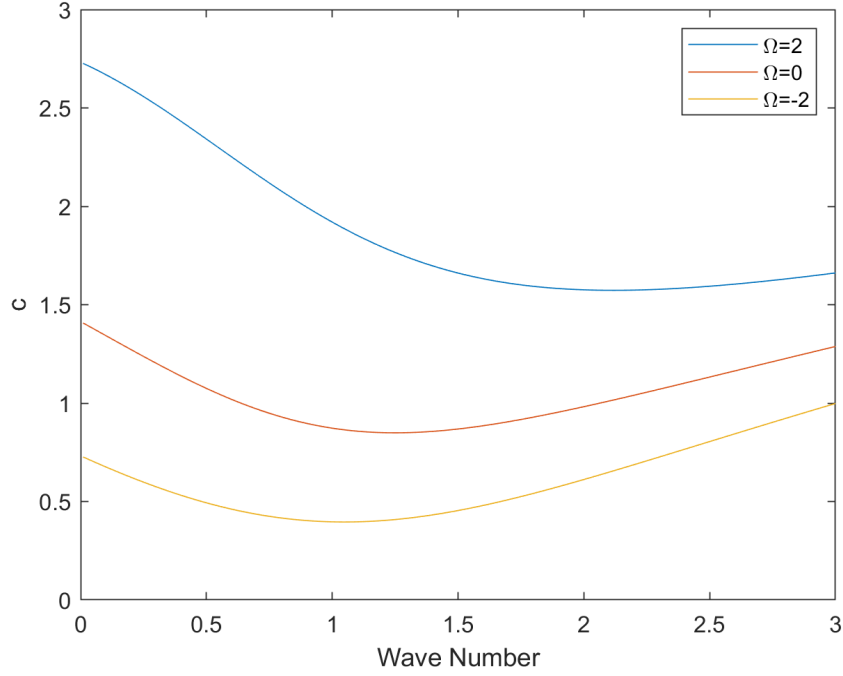}
\caption{\small\em Linear dispersion relation with constant vorticity for three fixed values of parameter $\Omega$.}
\label{fig2}
\end{center}
\end{figure}

\begin{figure}
\begin{center}
\includegraphics[width=0.8\textwidth]{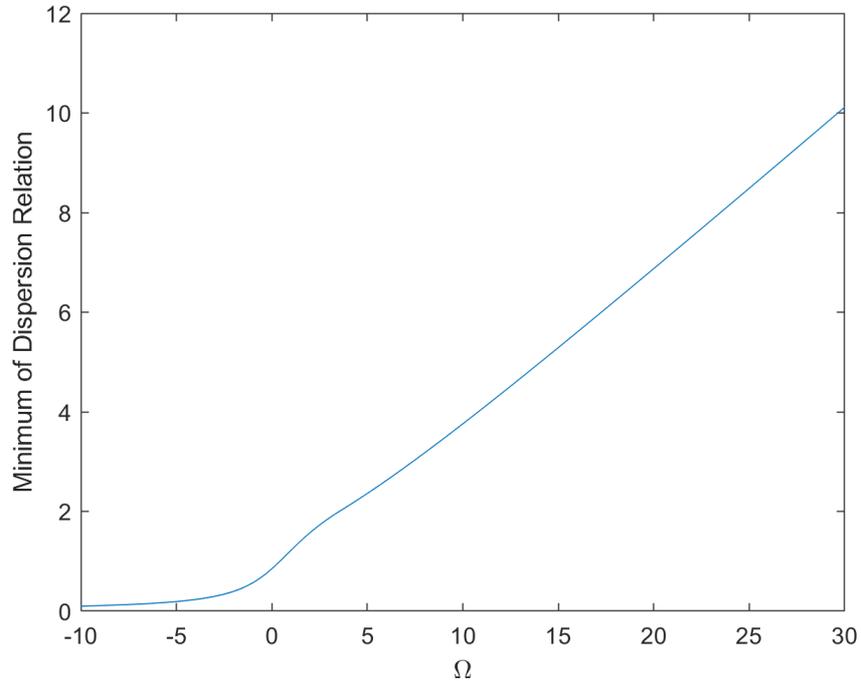}
\caption{\small\em Minimal value of the dispersion relation as a function of the vorticity parameter $\Omega$, with  $B=0.1$ and $E_{b}=\sqrt{0.2}$.}
\label{min_disp}
\end{center}
\end{figure}

As can be seen in Figure~\ref{fig2}, there is a minimum which is positive. For various choices of $B$ and $E_{b}$, there is a positive minimum for a wide range of vorticity(figure \ref{min_disp}). By differentiation of Equation~(\ref{disp_rel}), one can show:
\begin{equation}
\left.\frac{\ud c}{\ud k}\right\vert_{k=0}=-\frac{E_{b}}{\sqrt{\Omega^{2}+4B}}<0
\end{equation}
on the branch for which $c(0)>0$. One can also show that for large $k$, $c(k)\sim \sqrt{k}$ as $k\rightarrow\infty$ which shows the existence of a minimum in the dispersion relation. The beginning point at $k=0$ can be seen to satisfy the equation:
\begin{equation}
c^{2}-\Omega c-4B=0\Rightarrow c=\frac{\Omega\pm \sqrt{\Omega^{2}+4B}}{2}
\end{equation}

\subsection{Free Surface Profiles}

\begin{figure}
\centering
\subfigure[\label{fig:sub1}$\mu=0.015$, $U=0.55$]{\includegraphics[width=0.49\textwidth]{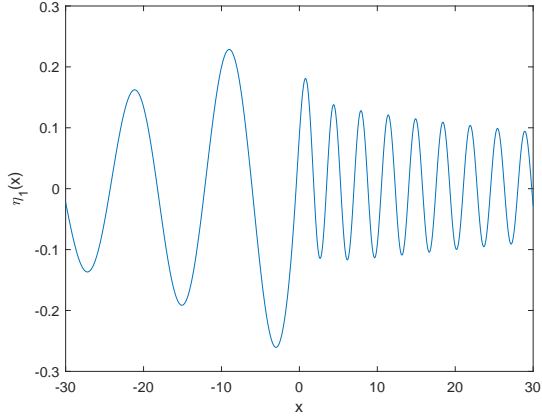}}
\subfigure[\label{fig:sub2}$\mu=0.0$, $U=0.7$]{\includegraphics[width=0.49\textwidth]{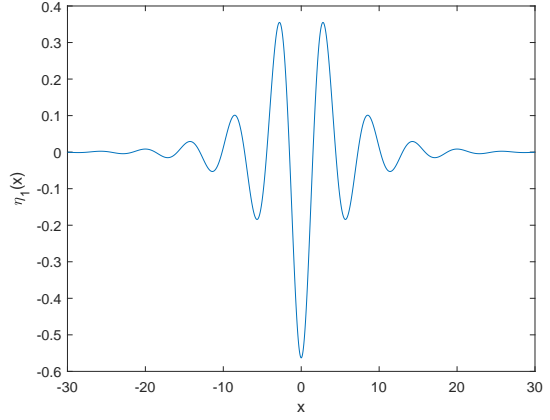}}
\caption{\small\em Free surface profiles under a moving pressure distribution.}
\end{figure}

Consider a moving pressure distribution moving with non-dimensional speed $U$. Then a frame of reference moving with speed $U$ is selected and all time derivative terms may be dropped. The horizontal velocity component is expanded as:
\begin{equation}
\hat{u}=U-\Omega y+\varepsilon u_{1}+o(\varepsilon )
\end{equation}
The equations which are changed are then:
\begin{eqnarray}
(U-\Omega y)\poo{u_{1}}{x} -\Omega v_{1} & = & -\poo{p_{1}}{x} \\
(U-\Omega y)\poo{v_{1}}{x} & = & -\poo{p_{1}}{y} \\
U\poo{\eta_{1}}{x} & = & \eta_{1}
\end{eqnarray}  
The method of derivation is very similar to that of the derivation of the dispersion relation will be omitted. The perturbations will be expressed at:
\begin{equation}
f(x,y)=\frac{1}{2\pi}\int_{\mathbb{R}}\hat{f}(k,y)e^{ikx}dk.
\end{equation}
The free surface is given by:
\begin{equation}\label{above}
\hat{\eta}=\frac{\hat{\mathcal{P}}_{1}\tanh k}{k\hat{U}^{2}+(-B+\Omega\hat{U}+E_{b}|k|-k^{2})\tanh k}
\end{equation}
Choosing a $U$ below the minimum of dispersion relation, one uses equation (\ref{above}). In order to compute the waves for which $U$ is above the minimum of the dispersion, one must include Rayleigh viscosity $\mu$ in the following way:
\begin{equation}
\hat{\eta}=\frac{\hat{\mathcal{P}}_{1}\tanh k}{k\hat{U}^{2}+(-B+\Omega\hat{U}+E_{b}|k|-k^{2})\tanh k+\mu i}   
\end{equation}
For $0<\mu<1$ gives waves of the form \ref{fig:sub1}. Figures \ref{fig:sub1} and \ref{fig:sub2} show the free surface profiles under a moving pressure distribution for $B\ =\ E_{b}\ =\ 2$ and $\Omega\ =\ 1\,$.

\section{The Weakly Nonlinear Free Surface}
\label{sec:wn}

To obtain a weakly nonlinear model of the phenomena, scale according to:
\begin{equation}
x=\lambda\hat{x},\quad y^{(1)}=h\hat{y},\quad y^{(2)}=\lambda\hat{y},\quad t=\frac{\lambda}{c_{0}}\hat{t},\quad u=c_{0}\hat{u},\quad v=\frac{hc_{0}}{\lambda}\hat{v}
\end{equation}
\begin{equation}
\eta =a\hat{\eta},\quad V=\lambda E_{0}\hat{V},\quad P=P_{a}-\rho gy-\frac{\epsilon_{d}E_{0}^{2}}{2}+\rho gh\hat{p}
\end{equation}
where $c_{0}=\sqrt{gh}$. The scaled equations are then:
\begin{eqnarray*}
\po{\hat{V}}{\hat{x}}{2}+\po{\hat{V}}{\hat{y}}{2} & = & 0 \\
\poo{\hat{V}}{x}+\alpha\sqrt{\beta}\poo{\hat{\eta}}{\hat{x}}\poo{\hat{V}}{\hat{y}} & = & 0 \\
\poo{\hat{u}}{\hat{t}}+\hat{u}\poo{\hat{u}}{\hat{x}}+\hat{v}\poo{\hat{u}}{\hat{y}} & = & -\poo{\hat{p}}{\hat{x}} \\
\beta\left(\poo{\hat{v}}{\hat{t}}+\hat{u}\poo{\hat{v}}{\hat{x}}+\hat{v}\poo{\hat{v}}{\hat{y}}\right) & = & -\poo{\hat{p}}{\hat{y}} \\
\beta\poo{\hat{v}}{\hat{x}}-\poo{\hat{u}}{\hat{y}} & = & \Omega \\
\poo{\hat{u}}{\hat{x}}+\poo{\hat{v}}{\hat{y}} & = & 0 \\
\poo{\hat{\eta}}{\hat{t}}+\hat{u}\poo{\hat{\eta}}{\hat{x}} & = &  \frac{\hat{v}}{\varepsilon} 
\end{eqnarray*}
Where $\Omega =h\omega /c_{0}$. The Young-Laplace equation becomes:
\begin{multline}
\hat{p}-\alpha\hat{\eta}-\frac{F_{E}^{2}}{2}=\alpha\hat{\mathcal{P}}-\frac{F_{E}^{2}}{1+\alpha^{2}\beta (\p_{\hat{x}}\hat{\eta})^{2}}(\alpha^{2}\beta (\p_{\hat{x}}\hat{\eta})^{2}\hat{T}_{11}-2\alpha\sqrt{\beta}\p_{\hat{x}}\hat{\eta}\hat{T}_{12}+\hat{T}_{22})-\\
-B\alpha\beta\frac{\p_{\hat{x}}^{2}\hat{\eta}}{(1+\alpha^{2}\beta (\p_{\hat{x}}\hat{\eta})^{2})^{\frac{3}{2}}}
\end{multline}
Where:
\begin{equation}
F_{E}^{2}=\frac{\epsilon_{d}E_{0}^{2}}{\rho gh},\quad B=\frac{\sigma}{\rho gh^{2}}
\end{equation}
The term $F_{E}$, is the ratio of a velocity to $c_{0}$ which shows that there is a natural velocity occurring which is given by, $U=\sqrt{\epsilon_{d}E_{0}^{2}/\rho}$, for this reason $F_{E}$ should be referred to as the electric Froude number. The next step is to make the transformation:
\begin{equation}
(\hat{p},\hat{u},\hat{v})=(\alpha\bar{p},-\Omega y+\alpha\bar{u},\alpha\bar{v})
\end{equation} 
The KdV scaling is $\alpha =\beta =\varepsilon$. The speed of propagation, $c$ of the (linear) waves is unknown at this point; thus the following co-ordinate transformation is used:
\begin{equation}
X=\hat{x}-c\hat{t},\quad T=\varepsilon \hat{t}
\end{equation}
Dropping bars and hats the equations then become:
\begin{eqnarray}
\po{V}{X}{2}+\po{V}{y}{2} & = & 0,\quad y>\varepsilon\eta \\
\poo{V}{X} +\varepsilon^{\frac{3}{2}}\poo{\eta}{X}\poo{V}{y} & = & 0,\quad  y=\varepsilon\eta 
\\
-c\poo{u}{X}+\varepsilon\poo{u}{T}+\varepsilon u\poo{u}{X}+\varepsilon v\poo{u}{y} & = & -\poo{p}{X},\quad -1<y<\varepsilon\eta \\
\varepsilon\left( -c\poo{v}{X}+\varepsilon\poo{v}{T}+\varepsilon u\poo{v}{X}+\varepsilon v\poo{v}{y}\right) & = & -\poo{p}{y},\quad -1<y<\varepsilon\eta \\
-c\poo{\eta}{X}+\varepsilon\poo{\eta}{T}+\varepsilon u\poo{\eta}{X} & = & v,\quad y=\varepsilon\eta \\
\varepsilon\poo{v}{X}-\poo{u}{y} & = & 0,\quad -1<y<\varepsilon\eta\label{wn1} \\
\poo{u}{X}+\poo{v}{y} & = & 0,\quad -1<y<\varepsilon\eta\label{wn2}
\end{eqnarray}
\begin{multline}
p-\eta -\frac{F_{E}^{2}}{2\varepsilon}=\mathcal{P}-\frac{1}{\varepsilon}\frac{F_{E}^{2}}{1+\varepsilon^{3}(\p_{X}\eta )^{2}}(\varepsilon^{2}(\p_{X}\eta )^{2}T_{11}-2\varepsilon^{\frac{3}{2}}\p_{X}\eta T_{12}-T_{22})-\\
-B\varepsilon\frac{\p_{X}^{2}\eta}{(1+\varepsilon^{3}(\p_{X}\eta )^{2})^{\frac{3}{2}}},\quad y=\varepsilon\eta
\end{multline}
It can be notes that the combination of equations (\ref{wn1}) and (\ref{wn2}) can be combined into:
\begin{equation}
\varepsilon\po{v}{X}{2}+\po{v}{y}{2}=0,\quad -1<y<\varepsilon\eta
\end{equation}
Expand according to:
\begin{eqnarray}
u(T,X,y) & = & -\frac{\Omega y}{\varepsilon}+u_{0}(T,X,y)+\varepsilon u_{1}(T,X,y)+o(\varepsilon )\quad y>\varepsilon\eta \\
v(T,X,y) & = & v_{0}(T,X,y)+\varepsilon v_{1}(T,X,y)+o(\varepsilon ) \\
p & = & p_{0}(T,X,y)+\varepsilon p_{1}(T,X,y)+o(\varepsilon ) \\
\mathcal{P} & = & \varepsilon\mathcal{P}_{1}+o(\varepsilon ) \\
V(T,X,y) & = & -y+\varepsilon^{\frac{3}{2}} V_{1}(T,X,y)+o(\varepsilon^{\frac{3}{2}}) \\
\eta & = & \eta_{0}(T,X)+\varepsilon\eta_{1}(T,X)+o(\varepsilon ) 
\end{eqnarray}
The $O(1)$ equations are:
\begin{equation}
\p_{y}p_{0}=0,\quad -(c+\Omega y)\p_{X}u_{0}-\Omega v_{0} =-\p_{X}p_{0},\quad \p_{y}^{2}v_{0}=0,\quad \p_{X}u_{0}+\p_{y}v_{0}
\end{equation}
With boundary conditions:
\begin{equation}
p_{0}=\eta_{0},\quad -c\p_{X}\eta_{0}=v_{0},\quad\textrm{on}\quad y=0
\end{equation}
The equation $\p_{y}^{2}v_{0}=0$ has solution:
\begin{equation}
v_{0}=(y+1)A_{0}(T,X)\Rightarrow \p_{X}u_{0}=-A_{0}(T,X)
\end{equation}
Setting $y=0$ shows that $A_{0}=-c\p_{X}\eta_{0}$ and so $u_{0}=c\eta_{0}$. The equation $\p_{y}p_{0}=0$ shows that $p_{0}=\eta_{0}$. The equation $-(c+\Omega y)\p_{X}u_{0}-\Omega v_{0} =-\p_{X}p_{0}$ can be evaluated at $y=0$. using the previous solutions yields:
\begin{equation}
-c^{2}\p_{X}\eta_{0}+\Omega c\p_{X}\eta_{0}=-\p_{X}\eta_{0}
\end{equation}
which gives the following expression for $c$,
\begin{equation}
c^{2}-\Omega c-1=0\Rightarrow c=\frac{\Omega\pm\sqrt{\Omega^{2} +4}}{2}
\end{equation}
The usual way to obtain this expression is to evaluate the Burns condition, which in this case is evaluating the integral:
\begin{equation}
\int_{-1}^{0}\frac{dz}{(c+\Omega z)^{2}}=1
\end{equation}
The method presented here bypasses the evaluation of increasingly complicated integrals with simple substitution. The next order equations are:
\begin{eqnarray}
\po{V_{1}}{X}{2}+\po{V_{1}}{y}{2} & = & 0 \\
V_{1}-\eta_{0} & = & 0 \\
-c\poo{u_{1}}{X}-\poo{u_{0}}{T}+u_{0}\poo{u_{0}}{X}-\Omega y\poo{u_{1}}{X}-\Omega v_{1} & = & -\poo{p_{1}}{X}\label{a1} \\
(c+\Omega y)\poo{v_{0}}{X} & = & -\poo{p_{1}}{y}\label{wn3} \\
\po{v_{0}}{X}{2}+\po{v_{1}}{y}{2} & = & 0\label{wn5} \\
-c\poo{\eta_{1}}{X}+\poo{\eta_{0}}{T}+(2c-\Omega )\eta_{0}\poo{\eta_{0}}{X} & = & v_{1}\label{wn6} \\
p_{1}-\eta_{1} & = & \mathcal{P}_{1}-\hat{F}_{E}^{2}\poo{V_{1}}{y}-B\po{\eta_{0}}{X}{2}\label{wn4}
\end{eqnarray}
Where the scaling on $F_{E}$ has been made, $F_{E}=\hat{F}_{E}\varepsilon^{1/4}$ to keep the electric term in the Young-Laplace equation. To progress, one finds the expression for $p_{1}$ by integrating (\ref{wn3}) and using (\ref{wn4}) to obtain:
\begin{equation}
p_{1}=c\left( \frac{(c-\Omega )}{2}[(y+1)^{2}-1]-\frac{\Omega}{3} [(y+1)^{3}-1]\right)\po{\eta_{0}}{X}{2}+\eta_{1}+\mathcal{P}_{1}-\hat{F}_{E}^{2}\poo{V_{1}}{y}-B\po{\eta_{0}}{X}{2}
\end{equation}
To obtain an equation, all one requires in $v_{1}(T,X,0)$. Solving equation (\ref{wn5}) for $v_{1}$ shows that:
\begin{equation}
v_{1}=\frac{c}{6}(y+1)^{3}\po{\eta_{0}}{X}{3}+A_{1}(T,X)(y+1)
\end{equation} 
Setting $y=0$ in equation (\ref{a1}) allows $A_{1}(T,X)$ to be computed:
\begin{multline}
(c-\Omega )A_{1}(T,X)=\left[ -B+\frac{\Omega c}{6}-\frac{c^{2}}{2}\right]\po{\eta_{0}}{X}{3}+\poo{u_{0}}{T}-u_{0}\poo{u_{0}}{X}-\poo{\eta_{1}}{X}-\poo{\mathcal{P}_{1}}{X}+\hat{F}_{E}^{2}\frac{\p^{2}V_{1}}{\p X\p y}
\end{multline}
Showing that:
\begin{multline}
A_{1}(T,X)=c\left[ B+\frac{\Omega c}{6}-\frac{c^{2}}{2}\right]\po{\eta_{0}}{X}{3}+c\poo{u_{0}}{T}-cu_{0}\poo{u_{0}}{X}-c\poo{\eta_{1}}{X}-c\poo{\mathcal{P}_{1}}{X}+c\hat{F}_{E}^{2}\frac{\p^{2}V_{1}}{\p X\p y}
\end{multline}
So this gives $v_{1}(T,X,0)$ to be:
\begin{multline}
v_{1}(T,X,0)=\frac{c}{6}\po{\eta_{0}}{X}{3}+c\left[ B+\frac{\Omega c}{6}-\frac{c^{2}}{2}\right]\po{\eta_{0}}{X}{3}-c\poo{u_{0}}{T}-cu_{0}\poo{u_{0}}{X}-c\poo{\eta_{1}}{X}-\\
-c\poo{\mathcal{P}_{1}}{X}+c\hat{F}_{E}^{2}\frac{\p^{2}V_{1}}{\p X\p y}
\end{multline}
To compute $V_{1}$, use the Fourier representation to obtain:
\begin{equation}
\po{\hat{V}_{1}}{y}{2}-k^{2}\hat{V}_{1}=0,\quad \hat{V}_{1}(t,k,0)=\hat{\eta}_{0}
\end{equation} 
to obtain:
\begin{equation}
V_{1}=\frac{1}{2\pi}\int_{\mathbb{R}}\hat{\eta}_{0}e^{-|k|y}e^{ikX}dk
\end{equation}
Then:
\begin{eqnarray*}
\poo{v_{1}}{y}(T,X,0) & = & \frac{1}{2\pi}\int_{\mathbb{R}}-|k|\hat{\eta}_{0}e^{ikX}dk \\
& = & \frac{1}{2\pi}\int_{\mathbb{R}}(ik)(i\textrm{sgn}(k))\hat{\eta}_{0}e^{ikX}dk \\
& = & \widehat{\mathscr{H}(\p_{X}\eta_{0})}
\end{eqnarray*}
Where $\mathscr{H}(\cdot )$ denotes the Hilbert transform. From the properties of Hilbert transforms:
\begin{equation}
\frac{\p^{2}V_{1}}{\p X\p y}=\mathscr{H}(\p_{X}^{2}\eta_{0})
\end{equation}
Inserting $v_{1}$ into equation (\ref{wn6}) yields the equation of interest:
\begin{equation}
(1+c^{2})\poo{\eta_{0}}{T}+(2c+c^{3}-\Omega)\eta_{0}\poo{\eta_{0}}{X}+c\left[\frac{c^{2}}{3}-B\right]\po{\eta_{0}}{X}{3}-c\hat{F}_{E}^{2}\mathscr{H}\left(\po{\eta_{0}}{X}{2}\right) +c\poo{\mathcal{P}}{X}=0
\end{equation}
A quick check by setting $c=1$ and $\Omega =0$ shows that it reduces to the original equation. Putting the dimensions back in shows that:
\begin{multline}
(1+c^{2})\left( c\poo{\eta}{x}+\frac{1}{c_{0}}\poo{\eta}{t}\right)+\frac{2c+c^{3}-\Omega}{h}\eta\poo{\eta}{x}+h^{2}c\left[\frac{c^{2}}{3}-B\right]\po{\eta}{x}{3} \\ -chF_{E}^{2}\mathscr{H}\left(\po{\eta}{x}{2}\right)+\frac{1}{\rho g}\poo{p}{x}=0
\end{multline}
The scaled pressure. Now travelling wave profiles are examined by $\eta =\eta (x-Vt)$ to get the equation:
\begin{equation}\label{eq:Benji}
(1+c^{2})(c-F)\eta +\frac{2c+c^{3}-\Omega}{h}\eta^{2}+h^{2}c\left[\frac{c^{2}}{3}-B\right]\eta ''-hcF_{E}^{2}\mathscr{H}(\eta')+\frac{p}{\rho g}=0
\end{equation}
Setting $\Omega=0$ which it follows that $c=1$ reduces to the equation in \cite{Hunt1}, in the paper \cite{Hur}, a weakly nonlinear equation for constant vorticity is derived comparable to the one with $E_{b}=0$.
\section{Numerical methods and results}
\label{sec:num}

In the sequel we consider the free wave propagation only, \emph{i.e.} $\tilde{p} \equiv 0$. In order to generate numerically the solitary wave solutions to Equation~\eqref{eq:Benji}, we employ the classical Fourier-type pseudo-spectral discretization on a sufficiently large domain, where the solution decays below the machine accuracy to annihilate the effect of implied periodic boundary conditions. In other words, we solve formally a periodic BVP, but the repeated value is actually zero in agreement with the decaying properties of solitary waves. The discrete problem for spectral coefficients is solved using the classical Petviashvili iteration as it was described in \cite{Dougalis2012} for the classical Benjamin equation. To implement the Petviashvili scheme, Equation~\eqref{eq:Benji} is written in the following form:
\begin{equation}\label{eq:tw}
    \mathcal{L}\eta\ =\ \mathcal{N}\,(\eta)\,,
\end{equation}
where the linear $\mathcal{L}$ and nonlinear $\mathcal{N}(\cdot)$ operators are defined as
\begin{align*}
  \mathcal{L}\eta\ &:=\ (1+c^{2})(c-F)\eta\ +\ h^{2}c\left[\frac{c^{2}}{3}-B\right]\eta ''\ -\ h\,c\,F_{\,E}^{2}\mathscr{H}(\eta'),\\
  \mathcal{N}\,(\eta)\ &:=\ \frac{\Omega\ -\ 2c\ -\ c^{3}}{h}\eta^{2}\,.
\end{align*}
Then, the Petviashvili iteration for Equation~\eqref{eq:tw} reads:
\begin{equation*}
    \eta^{(n+1)}\ =\ \gamma_n^{\,2}\,\mathcal{L}^{-1}\circ \mathcal{N}\,(\eta^{(n)})\,,
\end{equation*}
where we took into account the fact that the mapping $\mathcal{N}(\cdot)$ is a homogeneous function of degree two of its argument. Finally, the stabilizing factor $\gamma_n$ is defined as
\begin{equation*}
    \gamma_n\ :=\ \frac{\int_{\R} \eta_n\cdot \mathcal{L}\eta_n\: \ud x}{\int_{\R} \eta_n\cdot \mathcal{N}(\eta_n)\: \ud x} \,.
\end{equation*}
For more details on the Petviashvili iteration we refer to \cite{Dougalis2012, Alvarez2014}.

\subsection{Numerical results}

Equation~\eqref{eq:Benji} was discretized in space using the standard Fourier-type pseudo-spectral method. Namely, we used $N = 1024$ modes in our computations. Since we are looking at localized travelling waves, the computational domain was taken to be $T = [-10,\, 10]$ assuming periodic boundary conditions ensured by the choice of basis functions. The Petviashvili iterations were stopped when the $L_\infty$ norm of the difference of two successive iterations became smaller than $\epsilon = 5\times 10^{-15}$. As the initial guess we always took a localized bump of negative polarity. The convergence of the method was marginally dependent on our choice, which only influenced the total number of iterations. In any case, from the end user perspective, the computations lasted virtually instantaneously. We noticed that the number of iterations increased with the electric Froude number $F_E$. The computation of an oscillating travelling wave for $F_E = 1.27$ took $117$ Petviashvili iterations, while for $F_E = 0.5$ the method needed only $59$ iterations to converge. The dimensionless physical and numerical parameters used in this computation are reported in Table~\ref{tab:par}.

\begin{table}
    \centering
    \caption{\small\em Dimensional parameters used in numerical computations.}
    \smallskip
    \begin{tabular}{lc}
    \hline\hline
    \textit{Parameter name} & \textit{Value} \\
    \hline\hline
    Froude number, $F$ & $0.5$ \\
    Electric Froude number, $F_E$ & $0.5 \ldots 1.2$ \\
    Vorticity strength, $\Omega$ & $1.0$; $2.0$ \\
    Bond number, $B$ & $0.4$ \\
    Fluid layer depth, $d$ & $1.0$ \\
    Gravity acceleration, $g$ & $1.0$ \\
    Celerity, $c$ & $\frac12\bigl(\Omega - \sqrt{4 + \Omega^2}\bigr) \approx -0.62$ \\
    Number of Fourier modes, $N$ & $1024$ \\
    Domain half-length, $L$ & $10.0$ \\
    Error tolerance, $\epsilon$ & $5\times 10^{-15}$ \\
    \hline\hline
    \end{tabular}
    \label{tab:par}
\end{table}

\begin{figure}
    \centering
    \includegraphics[width=0.99\textwidth]{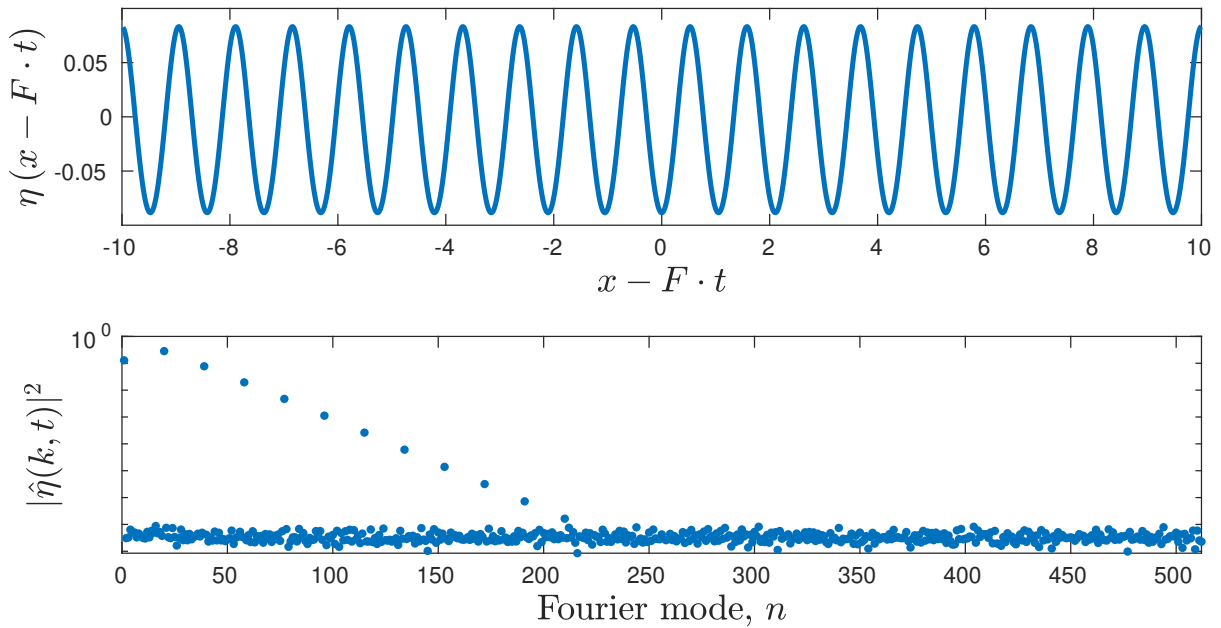}
    \caption{\small\em Fully converged travelling periodic wave solution to Equation~\eqref{eq:Benji} with $B = 0.1$ and $F_E = 0.5$. All other parameters are reported in Table~\ref{tab:par}. The bottom panel shows the Fourier power spectrum of the computed solution.}
    \label{fig:per}
\end{figure}

First of all, we would like to mention that for some values of parameters we were able to compute also the periodic travelling wave solutions, which was not our initial goal. One such solution is reported in Figure~\ref{fig:per}. Even if this solution appears to be a single Fourier mode (such as a sine wave), the Fourier power spectrum shown in the bottom panel of Figure~\ref{fig:per} shows that it is actually a superposition of infinite number of modes with exponentially decaying amplitude. The point is that only a finite number of them have an amplitude above the machine precision. Those were captured by the numerical method. We demonstrate this solution also because it is generic in some sense to Equation~\eqref{eq:Benji}. Namely, in our computations this type of solutions arose in a majority of the tested parameters values leading to the convergence of the algorithm.

\begin{figure}
  \centering
  \subfigure[$F_E = 0.5$]{\includegraphics[width=0.4\textwidth]{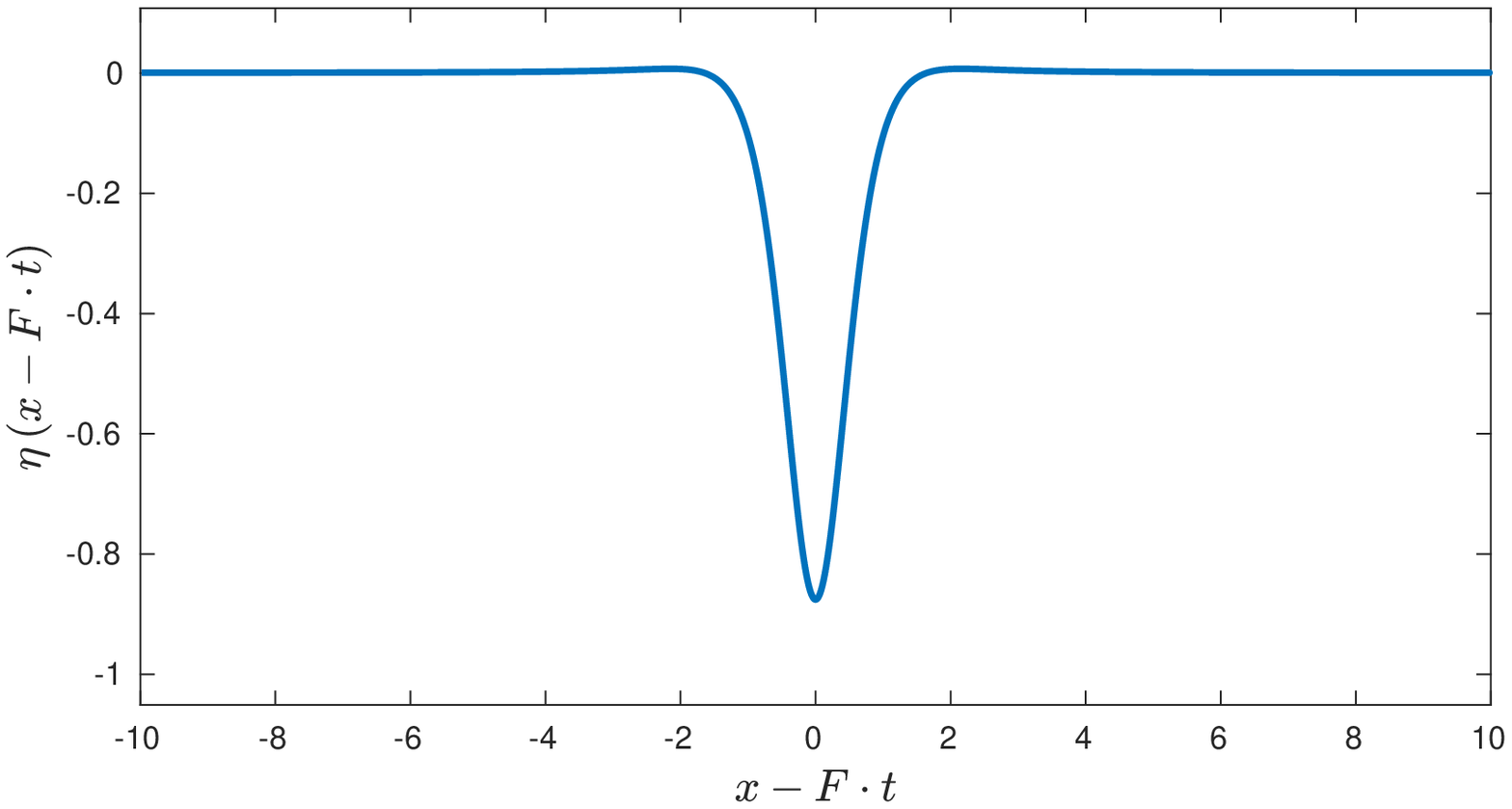}}
  \subfigure[$F_E = 0.9$]{\includegraphics[width=0.4\textwidth]{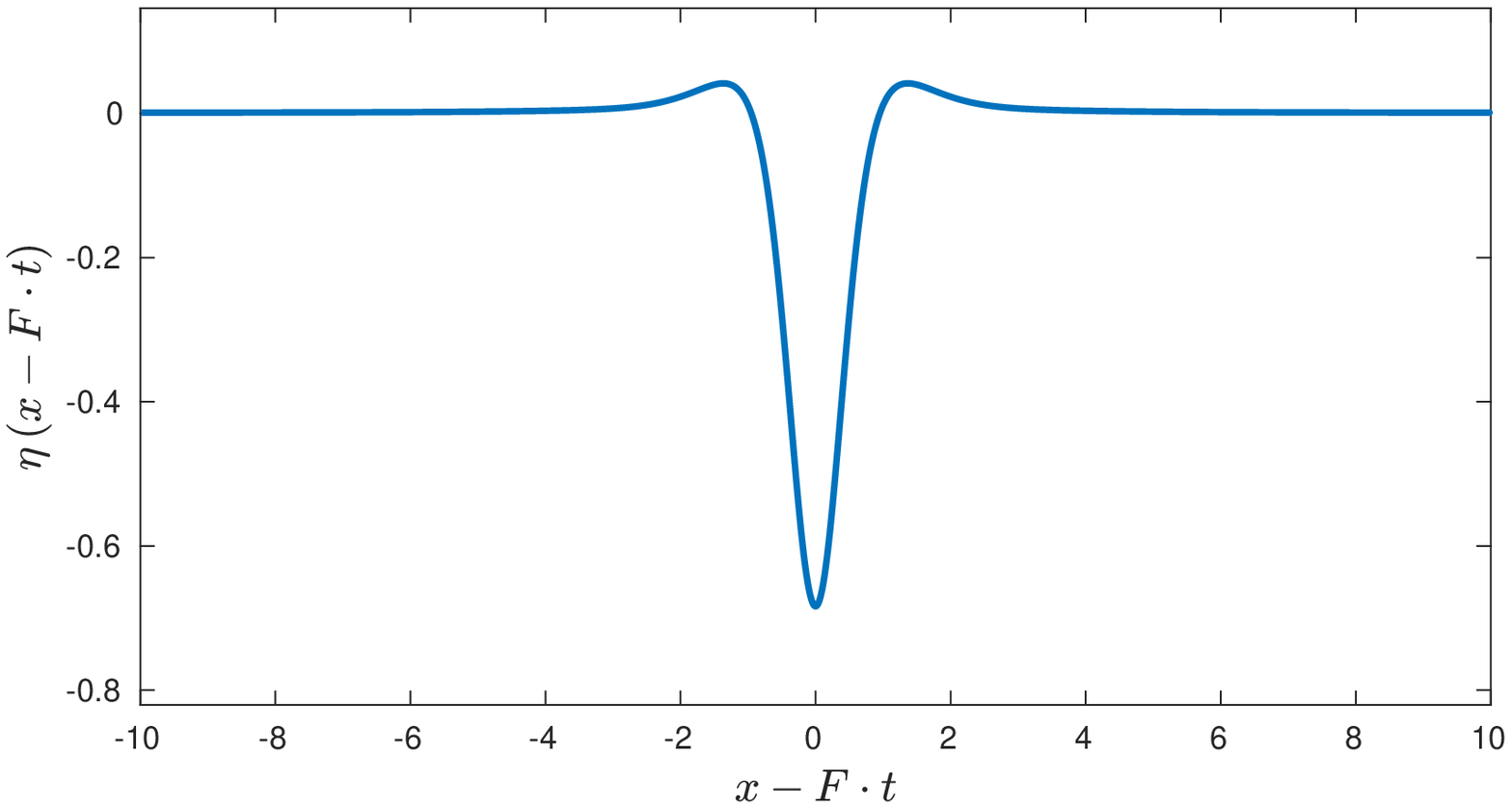}}
  \subfigure[$F_E = 1.22$]{\includegraphics[width=0.4\textwidth]{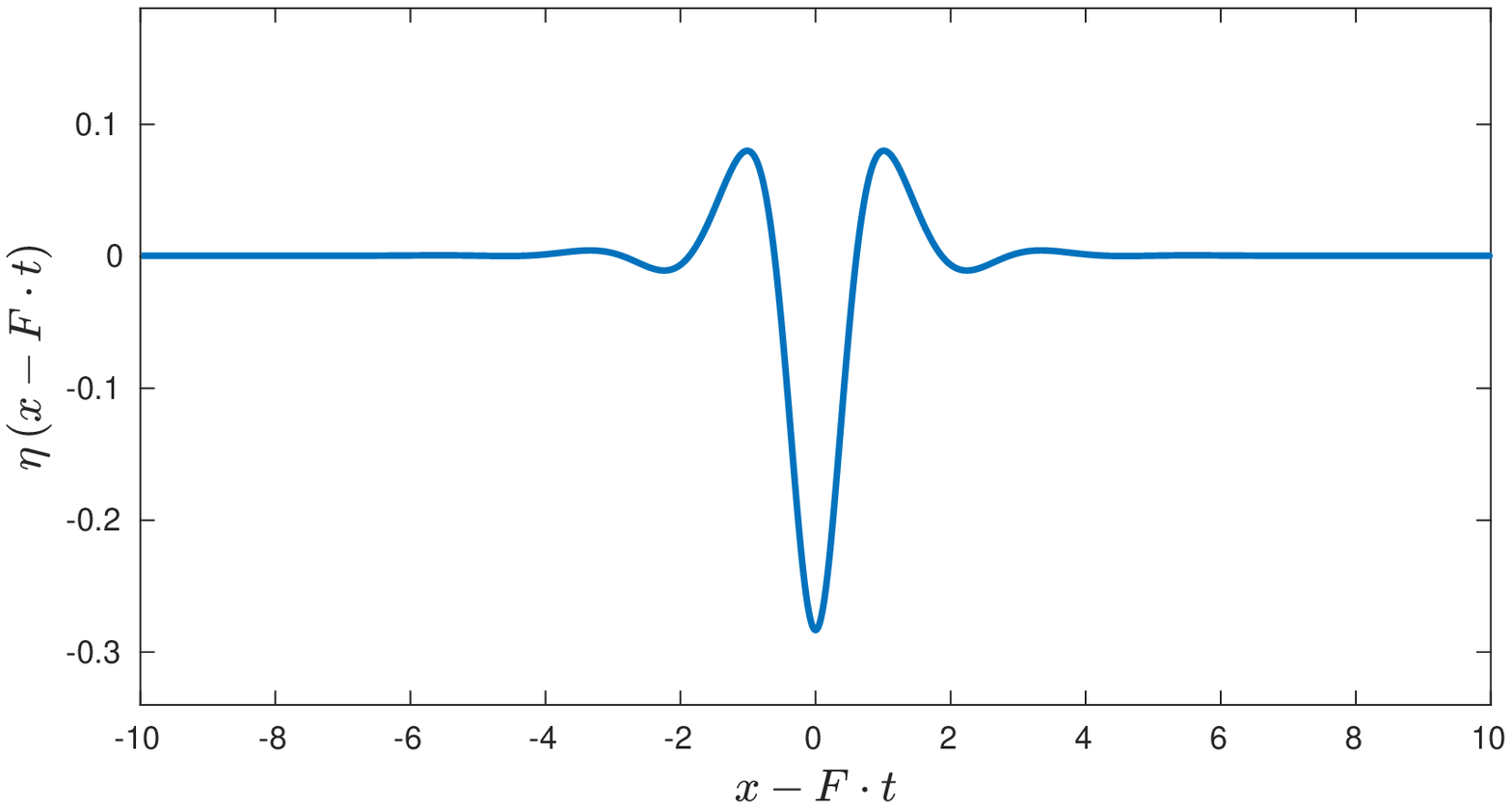}}
  \subfigure[$F_E = 1.27$]{\includegraphics[width=0.4\textwidth]{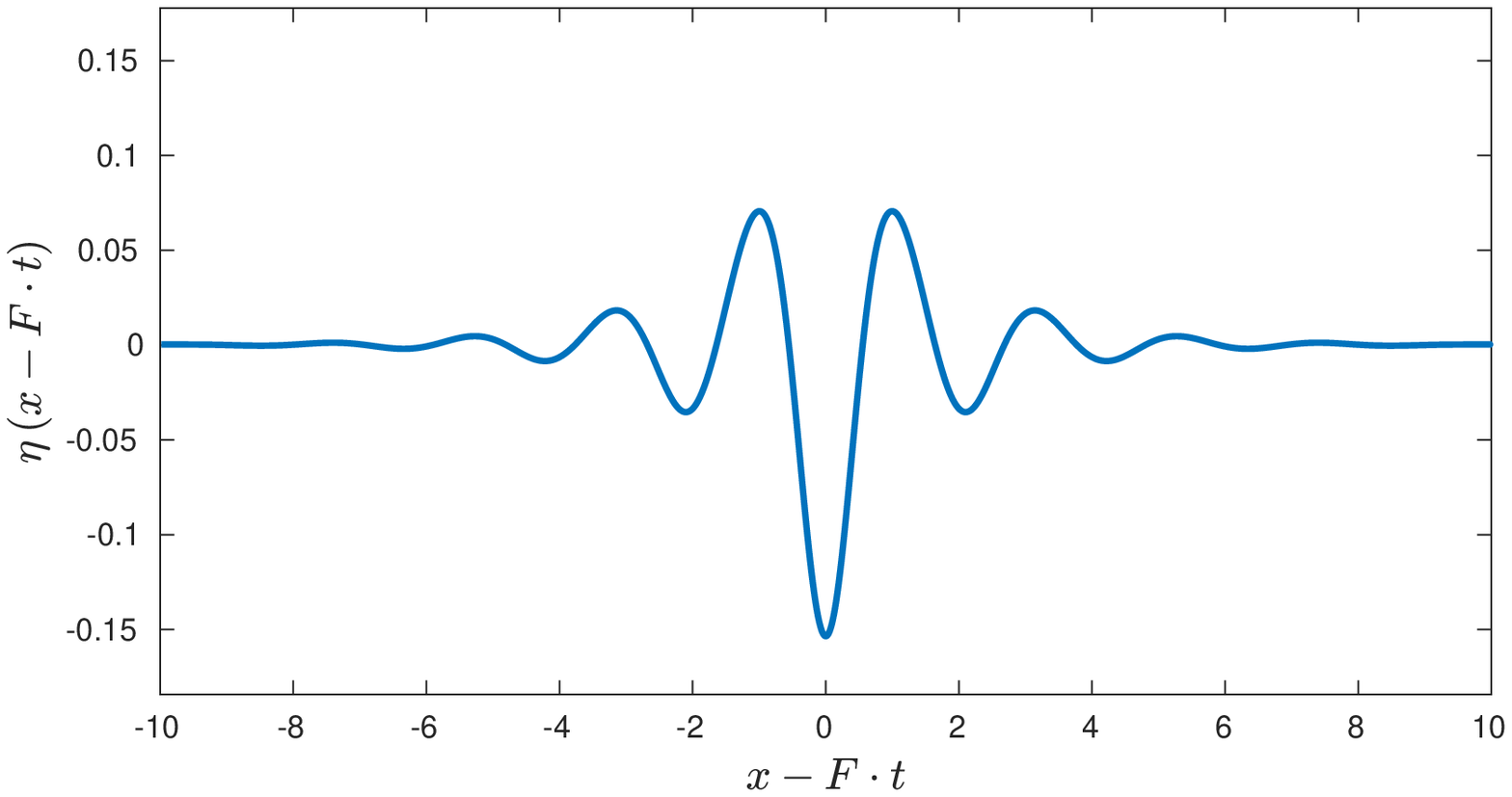}}
  \caption{\small\em Computed localized solitary waves to Equation~\eqref{eq:Benji} for the constant vorticity $\Omega\ =\ 1$ and different values of the electric Froude number $F_E$. All other parameters are reported in Table~\ref{tab:par}.}
  \label{fig:loc}
\end{figure}

\begin{figure}
  \centering
  \subfigure[$F_E = 0.5$]{\includegraphics[width=0.4\textwidth]{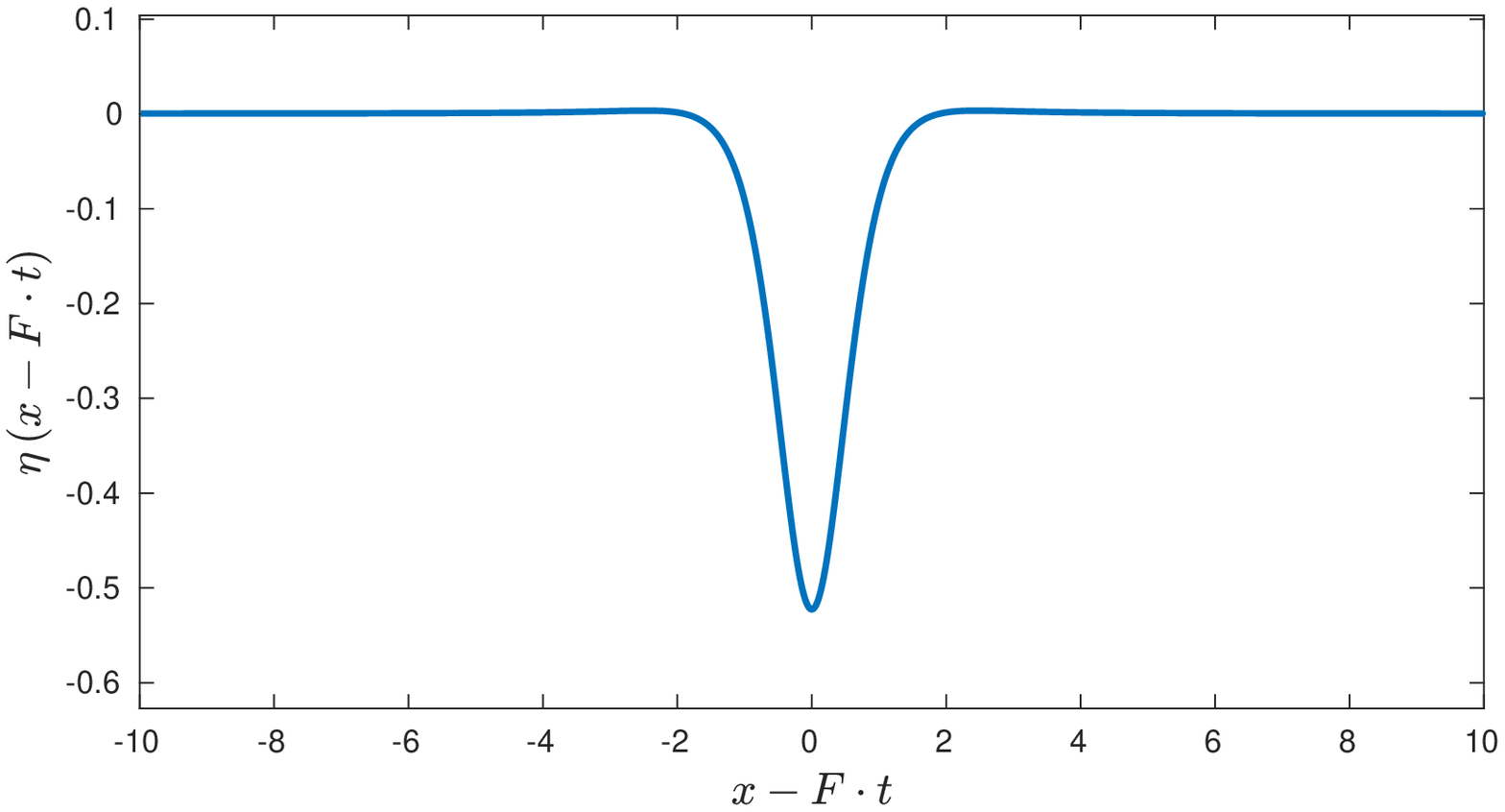}}
  \subfigure[$F_E = 0.9$]{\includegraphics[width=0.4\textwidth]{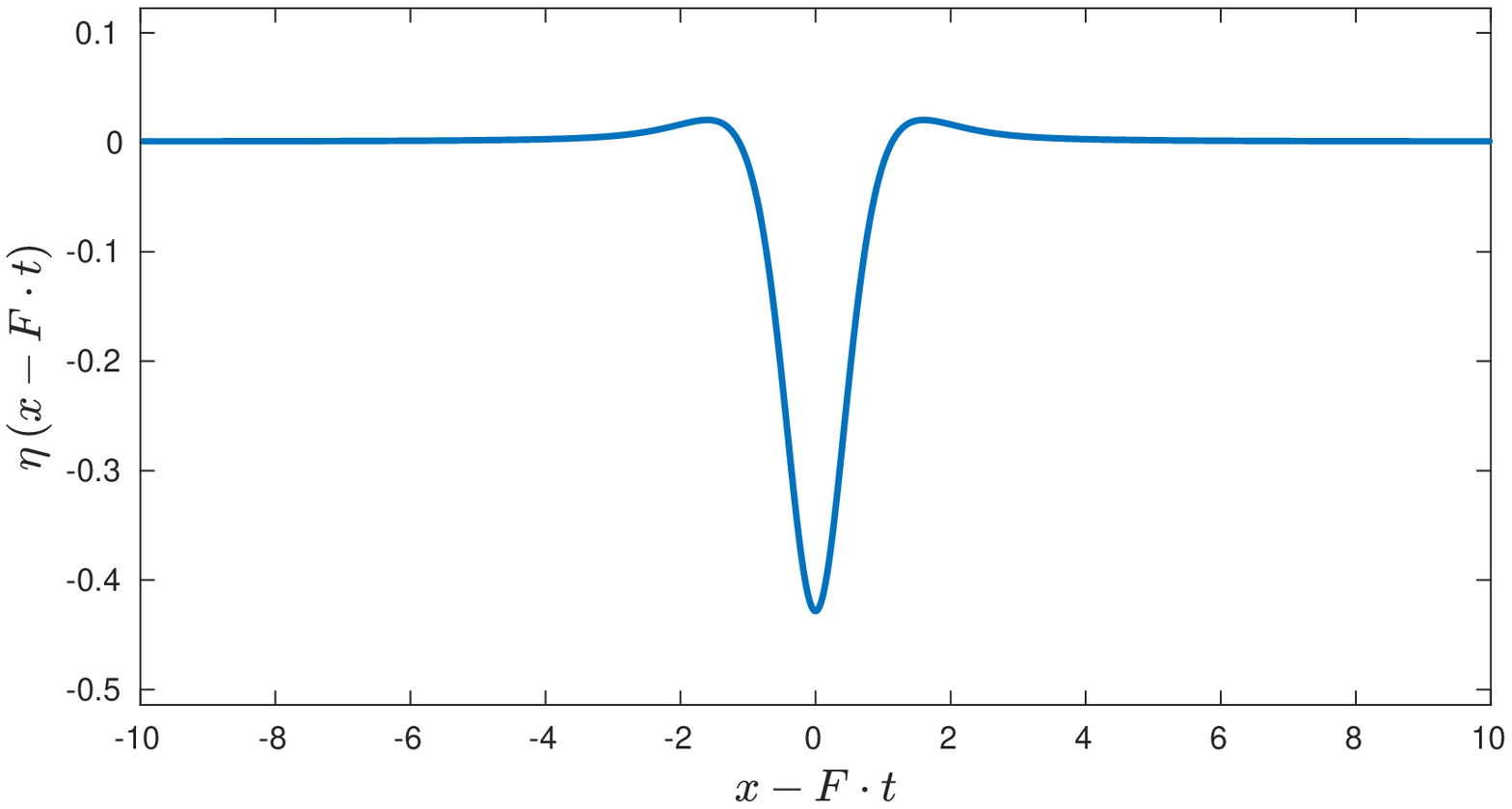}}
  \subfigure[$F_E = 1.22$]{\includegraphics[width=0.4\textwidth]{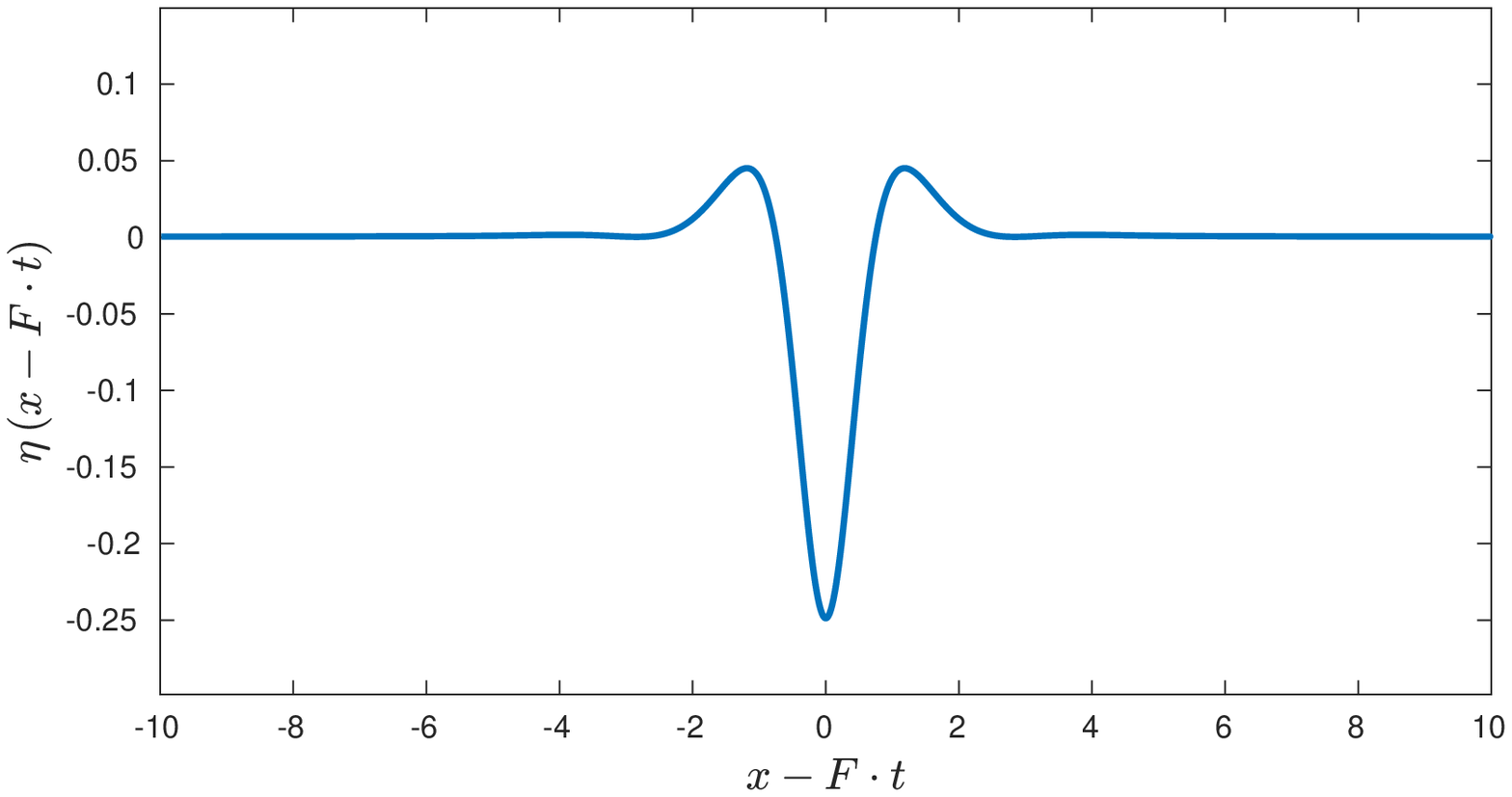}}
  \subfigure[$F_E = 1.27$]{\includegraphics[width=0.4\textwidth]{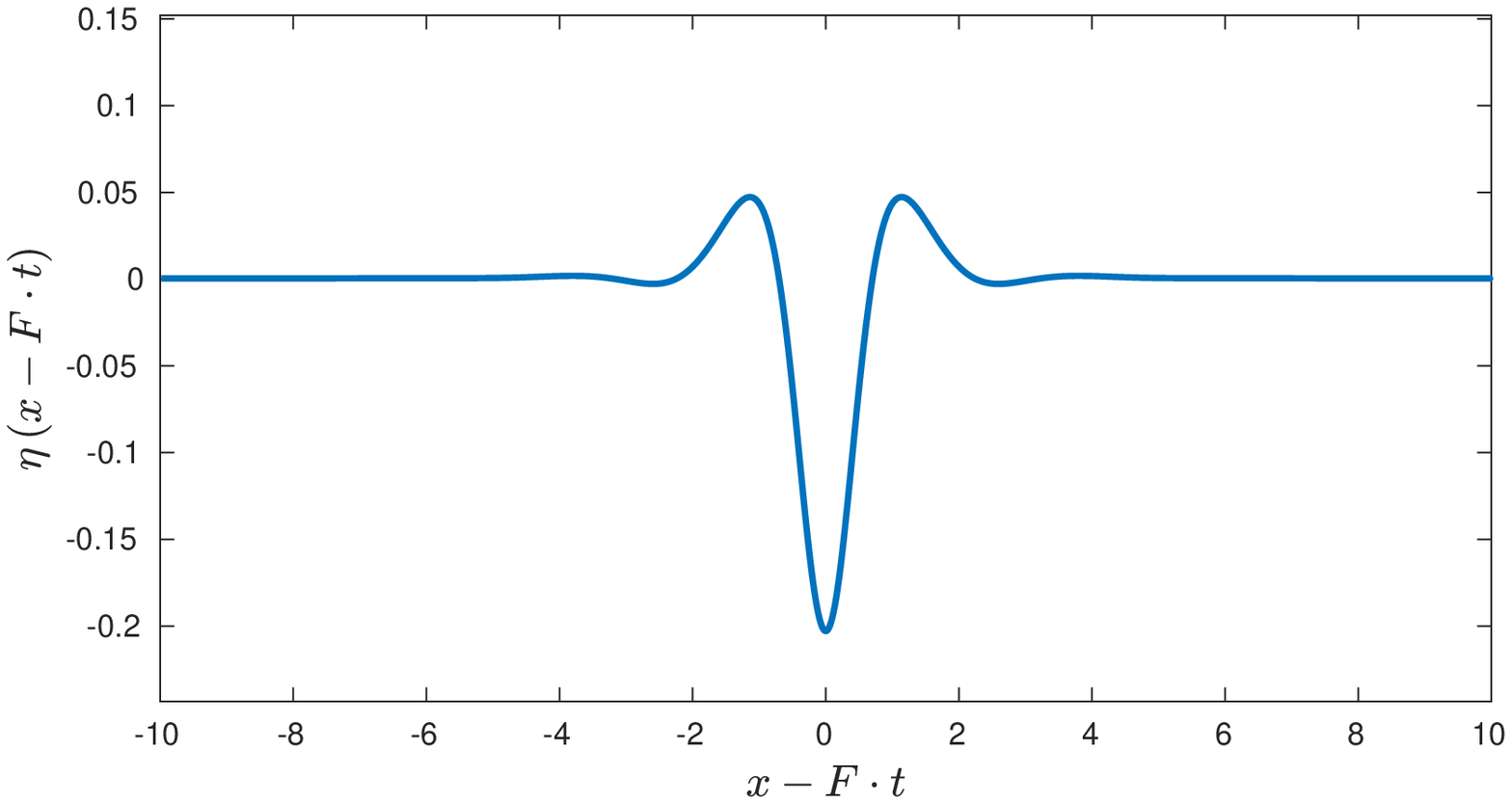}}
  \caption{\small\em Computed localized solitary waves to Equation~\eqref{eq:Benji} for the constant vorticity $\Omega\ =\ 2$ and different values of the electric Froude number $F_E$. All other parameters are reported in Table~\ref{tab:par}.}
  \label{fig:loc2}
\end{figure}

So, from now on we focus on localized (in space) travelling wave solutions, which are less generic but they are nevertheless present. Several examples of such structures are shown in Figure~\ref{fig:loc} (for $\Omega\ =\ 1$) and Figure~\ref{fig:loc2} (for $\Omega\ =\ 2$), where we gradually increased the electric Froude number $F_E$ by keeping all other values constant (see Table~\ref{tab:par}).

\begin{figure}
  \centering
  \subfigure[$F_E = 0.5$]{\includegraphics[width=0.48\textwidth]{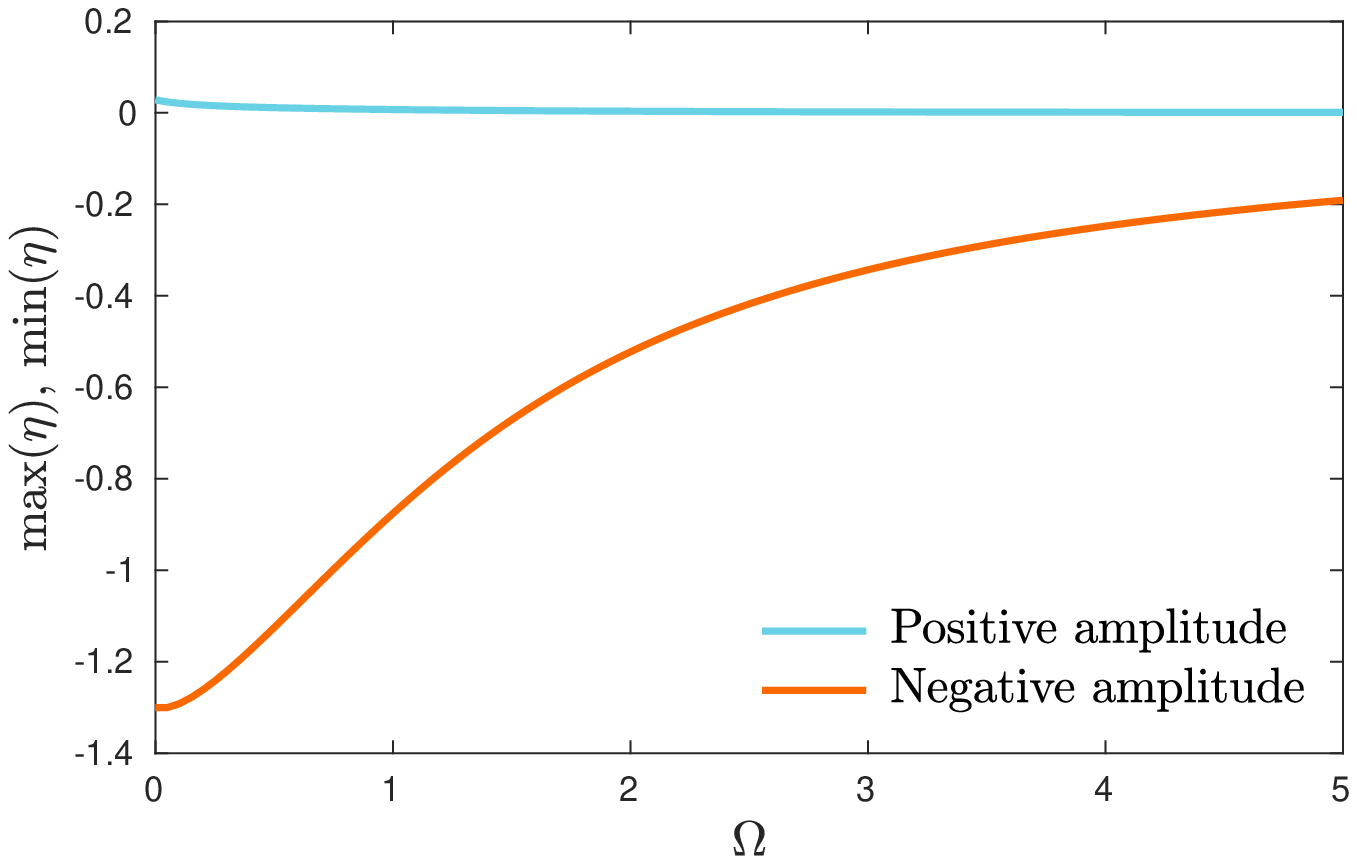}}
  \subfigure[$F_E = 0.9$]{\includegraphics[width=0.48\textwidth]{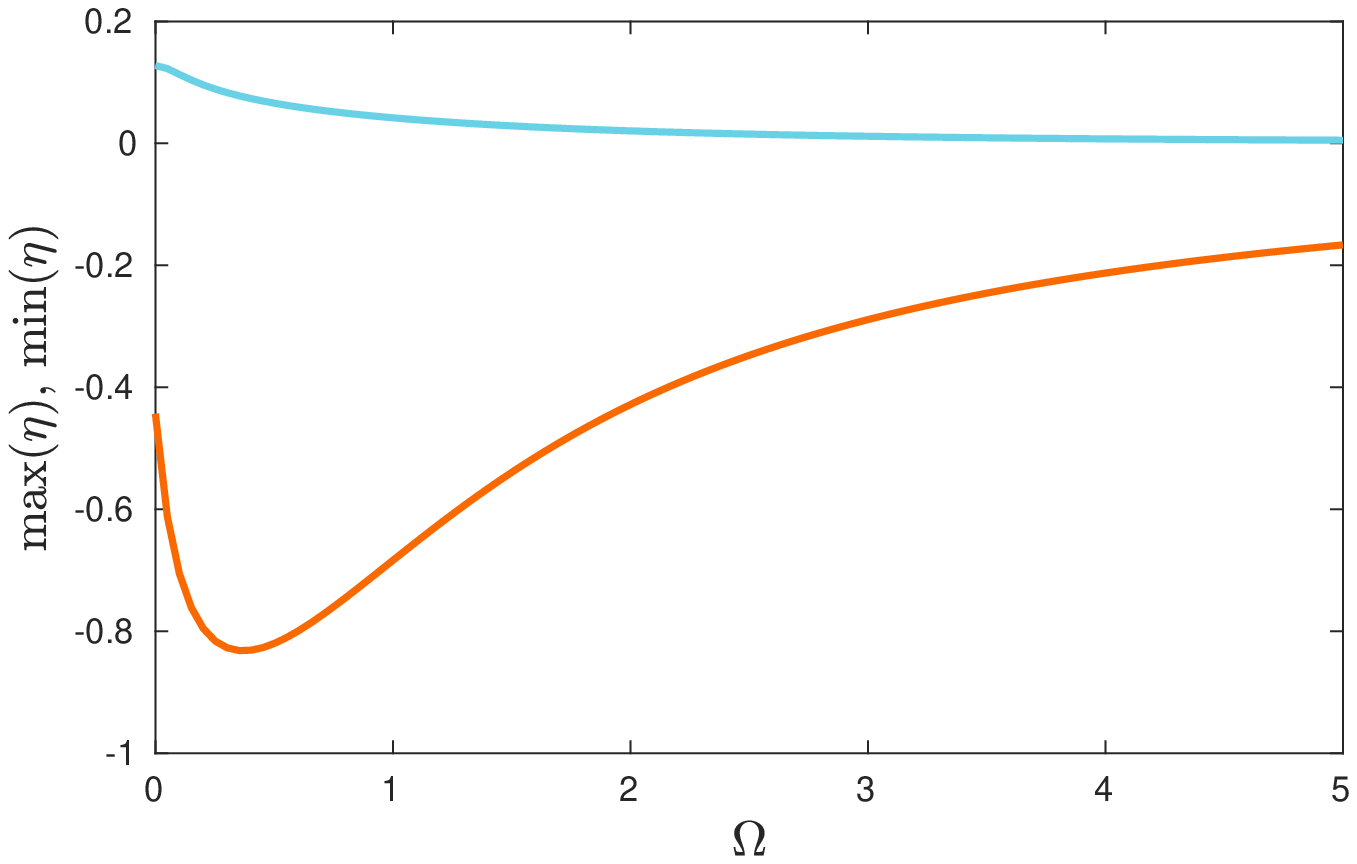}}
  \subfigure[$F_E = 1.22$]{\includegraphics[width=0.48\textwidth]{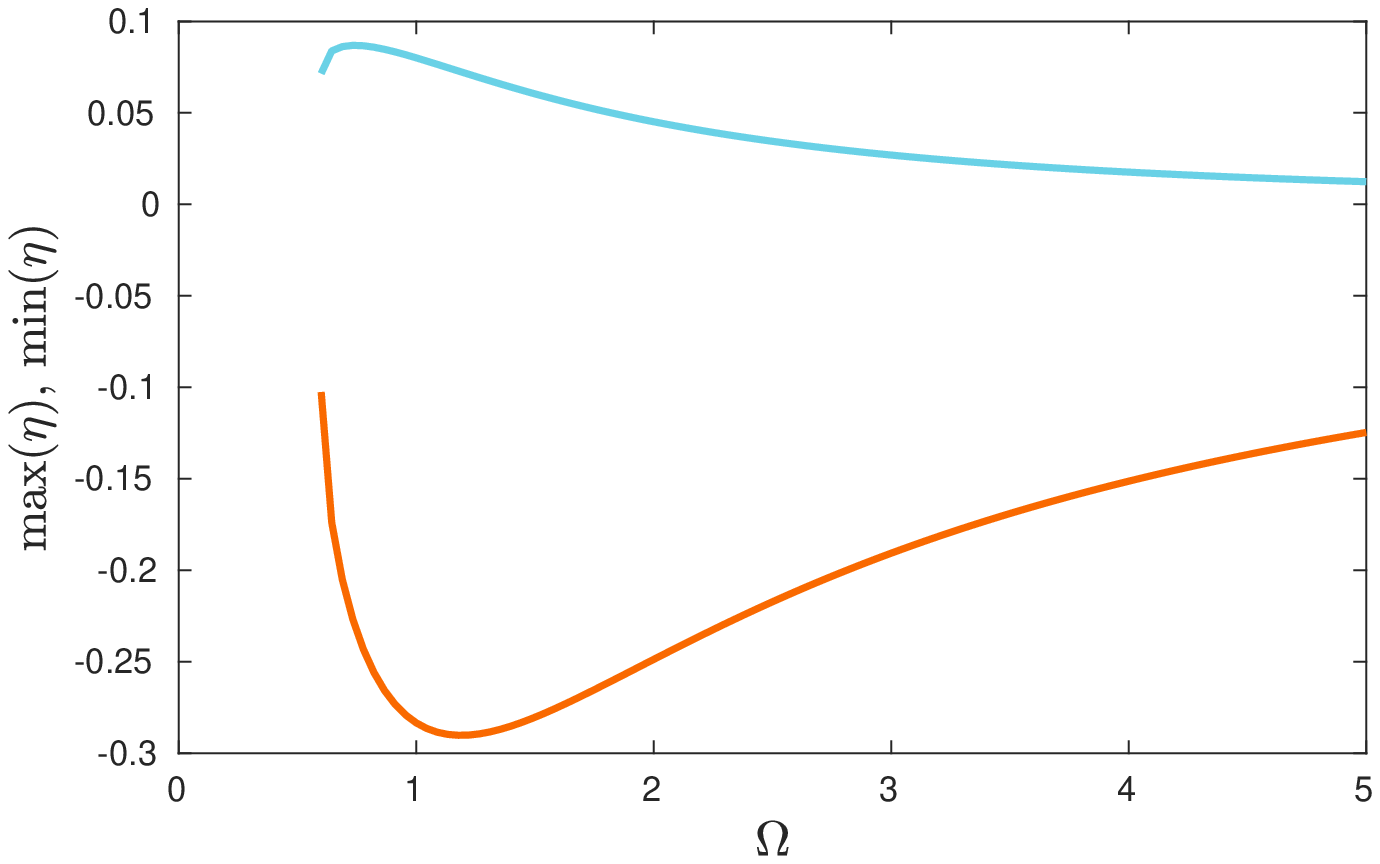}}
  \subfigure[$F_E = 1.27$]{\includegraphics[width=0.48\textwidth]{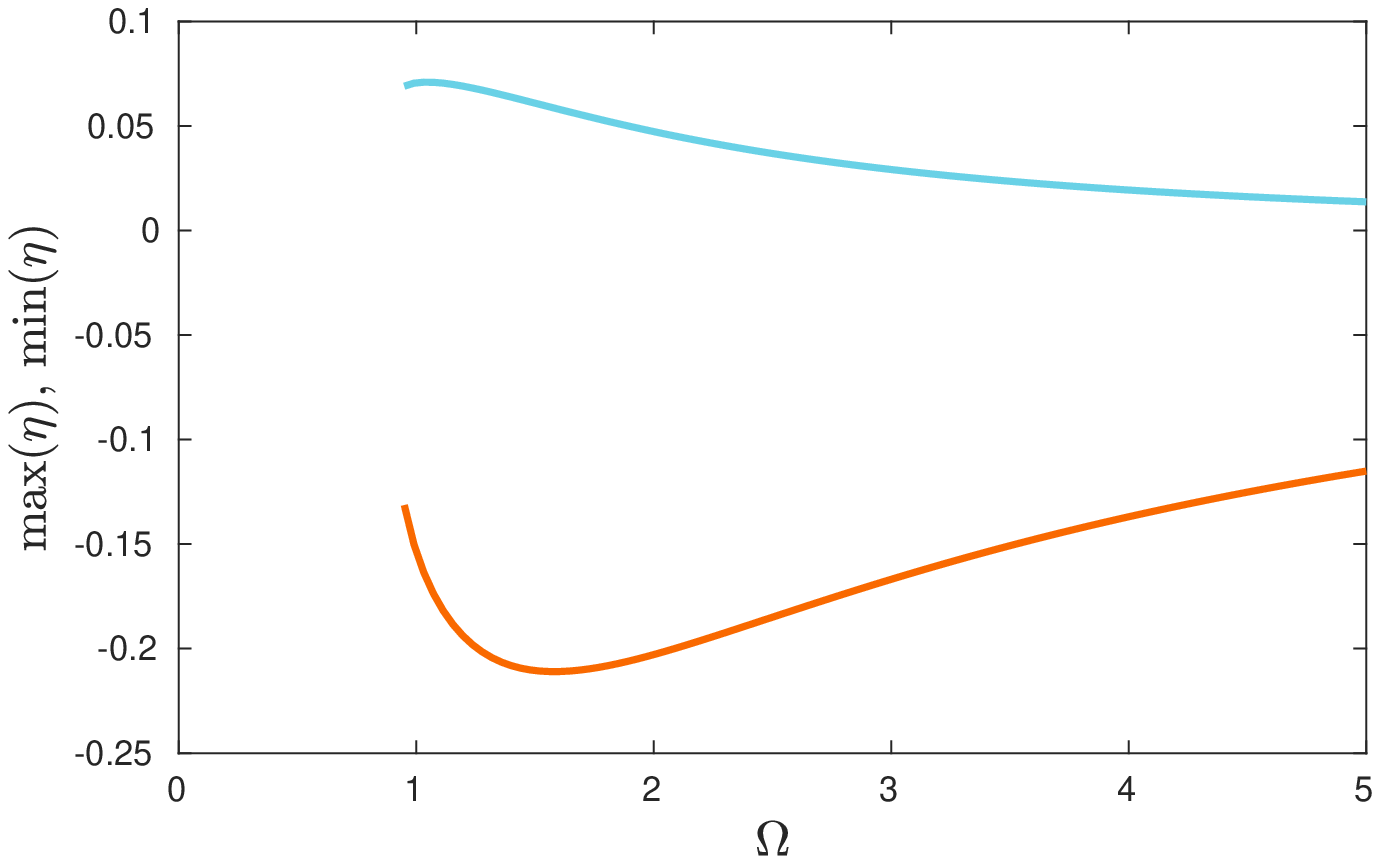}}
  \caption{\small\em The dependence of positive and negative travelling wave amplitudes on the vorticity parameter $\Omega$ for several values of the electric Froude number $F_E$. All other parameters (except $F_E$ and $\Omega$) are reported in Table~\ref{tab:par}.}
  \label{fig:ampls}
\end{figure}

\begin{figure}
  \centering
  \subfigure[$\Omega\ =\ 0.6$]{\includegraphics[width=0.48\textwidth]{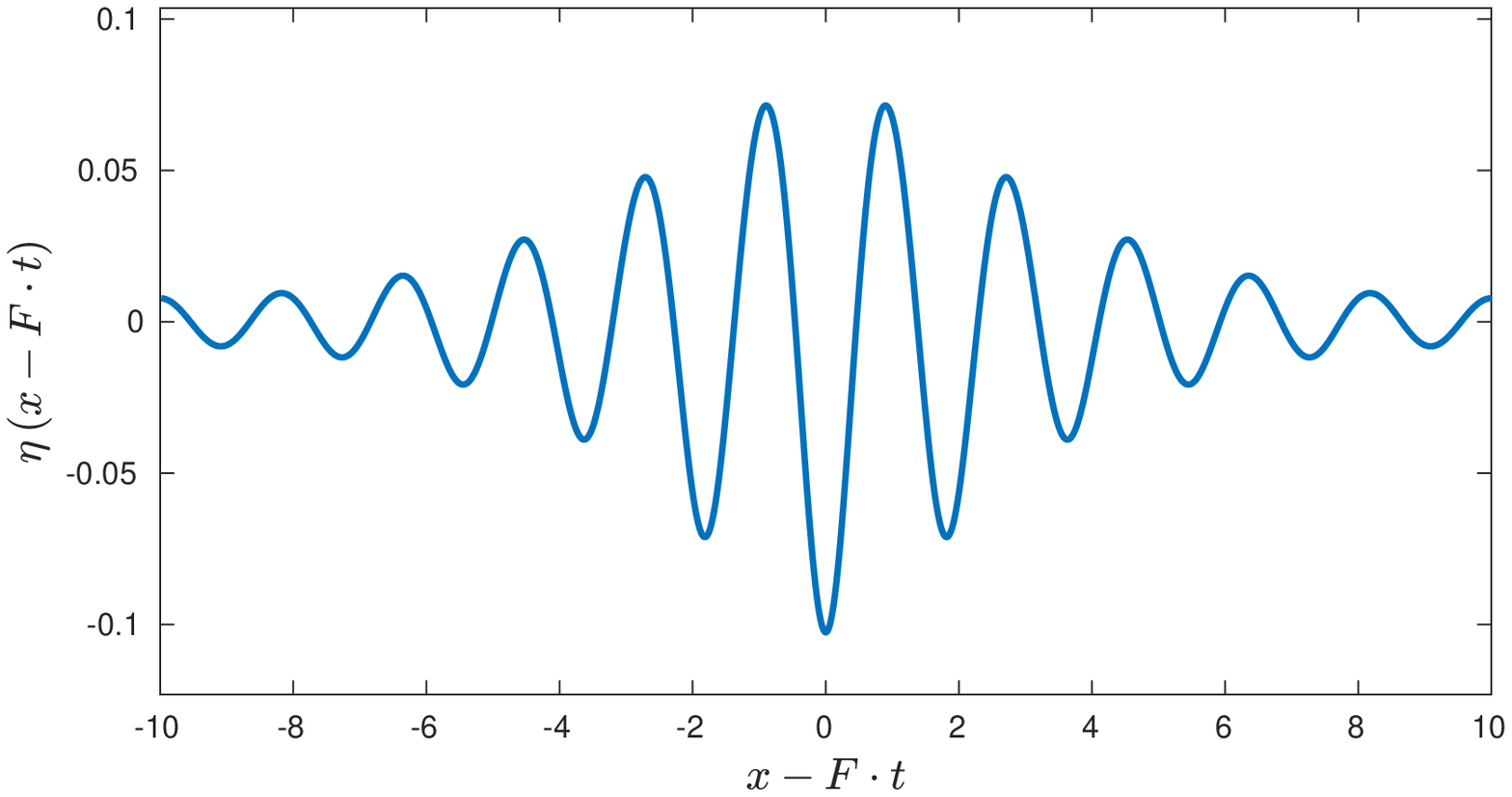}}
  \subfigure[$\Omega\ =\ 5.0$]{\includegraphics[width=0.48\textwidth]{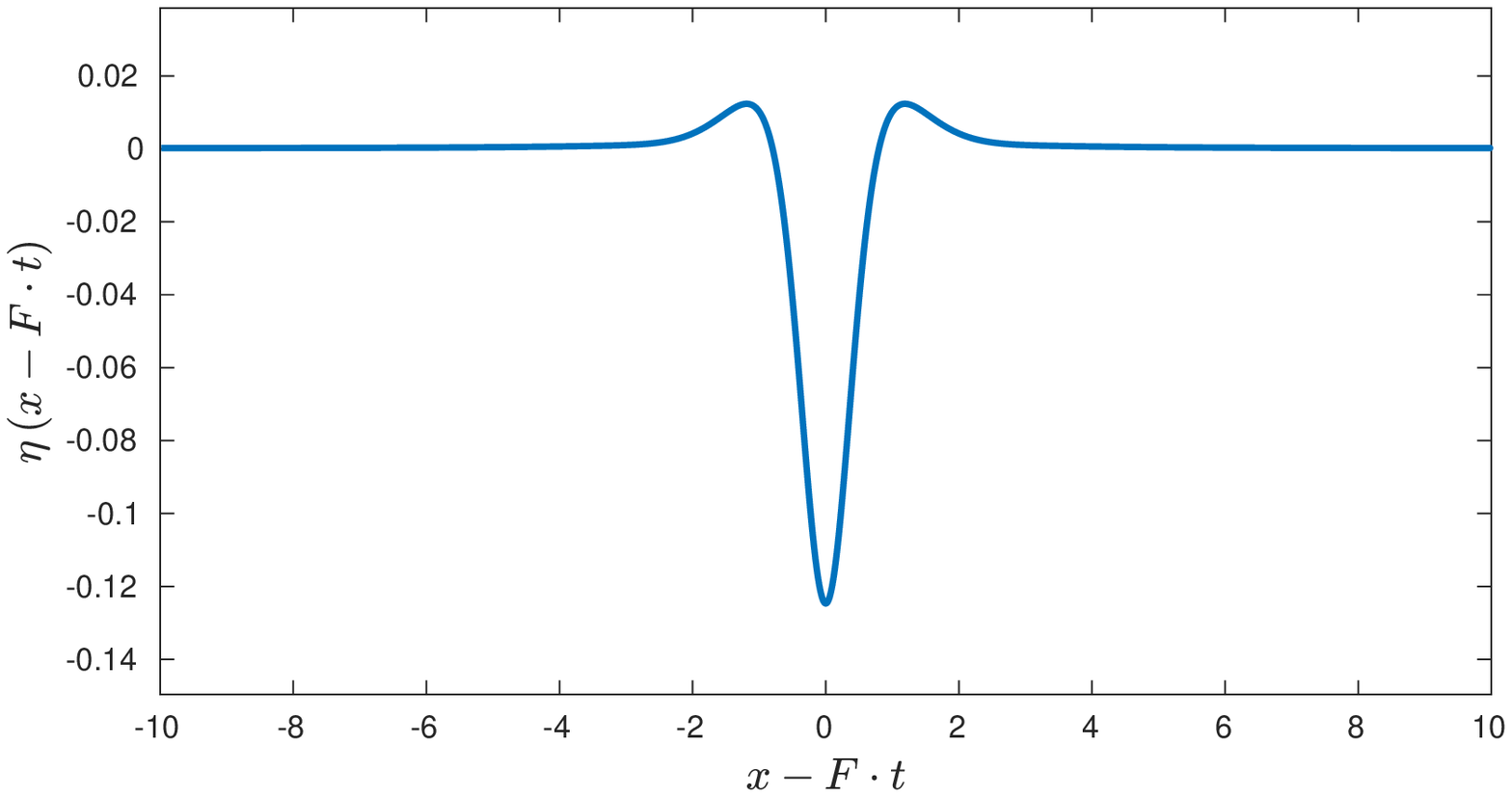}}
  \caption{\small\em Examples of a travelling wave solutions (for $F_E\ =\ 1.22$) near the lowest positive value of the vorticity parameter $\Omega$ (the left panel, here we took $\Omega\ =\ 0.6$) and the highest value taken in this study (the right panel, $\Omega\ =\ 5.0$). All other parameters (except $F_E$ and $\Omega$) are reported in Table~\ref{tab:par}.}
  \label{fig:minmax}
\end{figure}

We study also the dependence of the localized travelling wave solutions (such as depicted in Figures~\ref{fig:loc} and \ref{fig:loc2}) on the vorticity parameter $\Omega\,$. For this purpose we performed a series of computations with varying parameter $\Omega$ between its lowest positive value for which the travelling solution exists and the maximal value fixed here to be $\Omega\ =\ 5.0\,$. The result is presented in Figure~\ref{fig:ampls}. The dependence turns out to be non-monotonic. Though, for high values of the vorticity parameter, the positive amplitude seems to decrease and the negative amplitude seems to increase monotonically. To illustrate the shape of solutions at two extremes on Figure~\ref{fig:ampls}(\textit{c}), a typical solution near the lowest positive vorticity value is depicted in Figure~\ref{fig:minmax}(\textit{a}) while the solution corresponding to the highest value is depicted in Figure~\ref{fig:minmax}(\textit{b}).

\section{Conclusions and perspectives}
\label{sec:concl}

In the present manuscript we considered the propagation of free surface electrohydrodynamic waves in the presence of non-zero, but constant vorticity distribution. The problem was analyzed from the linear and weakly nonlinear points of view. The linear analysis allowed us to get rid of Burns's condition. The weakly nonlinear approach allowed us to compute solitary wave solutions by solving numerically the non-local ODE which describes them. The non-local effects are described by a linear term involving the Hilbert transform of the free surface excursion derivative $\eta^{\,\prime}$. It turns out that the dynamics of weakly nonlinear electrohydrodynamic waves is described by a generalized Benjamin equation, which appears clearly in this context for the first time, to our knowledge. So far, it appeared as a model equation for internal capillary-gravity waves. In our study it serves to predict the shape of coherent structures in electrohydrodynamic flows with constant vorticity.

Concerning the perspectives, in future works we would like to consider more general vorticity distributions. Another promising direction consists in considering the three-dimensional wave propagation problem. Finally, the unsteady simulations have to be performed to understand better the dynamics of solutions discussed hereinabove. We suspect also that the derived Benjamin-type Equation~\eqref{eq:Benji} possesses also other types of travelling wave solutions such as multi-pulsed solitary waves which were computed in \cite{Dougalis2012} in the context of internal waves.


\bibliography{biblio}
\bibliographystyle{abbrv}

\end{document}